\documentclass[aps,pra,twocolumn,superscriptaddress,nofootinbib,10pt]{revtex4-2}
\usepackage{amsmath}
\usepackage{amssymb}
\usepackage{hyperref}
\usepackage{graphicx}
\usepackage{bbm}
\usepackage{physics}
\usepackage{tikz}
\usetikzlibrary{3d, shadows, patterns, decorations.pathreplacing} 
\usepackage[svgnames]{xcolor}
\usepackage[T1]{fontenc}
\usepackage[normalem]{ulem}
\usepackage{mathtools}
\usepackage{yhmath}
\usepackage{mathbbol}
\usepackage{upgreek}
\usepackage{subfigure}
\usepackage{pgfplots}
\usepackage{bm} 
\usepackage{verbatim}

\begin{document}

\title{Improving Single Excitation Fidelity in Rydberg Superatoms \\ for Efficient Single Photon Emission}

\author{Vidisha Aggarwal}
\email{v.aggarwal@fz-juelich.de}
\affiliation{Forschungszentrum Jülich GmbH, Peter Grünberg Institute, Quantum Control (PGI-8), 52425 Jülich, Germany}
\affiliation{Institute for Theoretical Physics, University of Cologne, Zülpicher Straße 77, 50937 Cologne, Germany}

\author{Boxi Li}
\affiliation{Forschungszentrum Jülich GmbH, Peter Grünberg Institute, Quantum Control (PGI-8), 52425 Jülich, Germany}

\author{Eloisa Cuestas}
\affiliation{Forschungszentrum Jülich GmbH, Peter Grünberg Institute, Quantum Control (PGI-8), 52425 Jülich, Germany}
\affiliation{OIST Graduate University, Onna, Okinawa, Japan}

\author{Tommaso Calarco}
\affiliation{Forschungszentrum Jülich GmbH, Peter Grünberg Institute, Quantum Control (PGI-8), 52425 Jülich, Germany}
\affiliation{Institute for Theoretical Physics, University of Cologne, Zülpicher Straße 77, 50937 Cologne, Germany}
\affiliation{Dipartimento di Fisica e Astronomia, Università di Bologna, 40127 Bologna, Italy}

\author{Robert Zeier}
\affiliation{Forschungszentrum Jülich GmbH, Peter Grünberg Institute, Quantum Control (PGI-8), 52425 Jülich, Germany}

\author{Alexei Ourjoumtsev}
\affiliation{JEIP, UAR 3573 CNRS, Coll\`ege de France, PSL University 11, place Marcelin Berthelot, 75231 Paris Cedex 05, France}

\author{Felix Motzoi}
\affiliation{Forschungszentrum Jülich GmbH, Peter Grünberg Institute, Quantum Control (PGI-8), 52425 Jülich, Germany}
\affiliation{Institute for Theoretical Physics, University of Cologne, Zülpicher Straße 77, 50937 Cologne, Germany}

\date{\today}  

\begin{abstract}
Deterministic single photon emission from a Rydberg ensemble coupled to an optical cavity requires high-fidelity preparation of collective single excitations. In such a setup imperfect Rydberg blockade can lead to unwanted double excitations, which degrade photon indistinguishability. In this work we adapt the Derivative Removal by Adiabatic Gate (DRAG) technique, originally developed for superconducting qubits, to shape optical pulses that suppress double excitations in this atomic platform. By combining analytical modeling with numerical optimization, DRAG provides an improvement over conventional sine-squared pulses. Further optimization of pulse duration and atomic ensemble size identifies a parameter regime, distinct from that used in [Nature Photonics \textbf{17}, 688 (2023)], that enhances the single excitation probability from the previous theoretical benchmark of 77\% to 91.9\%, approaching the fundamental limits set by decoherence in the system. Benchmarking against GRAPE (Gradient Ascent Pulse Engineering) confirms that DRAG operates close to the optimal control limit, while maintaining smooth, experimentally feasible pulse shapes. These results demonstrate the effectiveness and cross platform adaptability of DRAG for a high-fidelity single photon source.
\end{abstract}

\maketitle

\section{Introduction}
\label{sec_intro}

The realization of large scale quantum communication networks, enabling secure quantum key distribution, distributed quantum computing, and other entanglement-based protocols~\cite{Gisin_2002_Rev_Mod_Phys,Gisin_2007_Nature_Photonics,kimble_2008_nature,Wehner_2018_Science}, relies on the efficient generation, manipulation, and transmission of quantum information over long distances. Optical photons are a natural choice for this purpose, as they exhibit minimal loss in fiber optic channels, propagate at relativistic speeds, and can be efficiently manipulated at room temperature. However, the inherent absence of photon-photon interactions at optical frequencies poses a fundamental challenge to their direct application in quantum systems. While linear optical elements can guide, interfere, and measure photons, they lack the intrinsic nonlinearity required for photon-photon interactions~\cite{Knill_2001_Nature}. To overcome this limitation, hybrid quantum architectures have emerged as a promising solution. These systems couple photons to matter-based platforms such as trapped atoms~\cite{Giannelli_2018_NJP,Morin_2019_PhysRevLett,Hacker_2019-NaturePhotonics}, ions~\cite{Schupp_2021_PRXQuantum,Ward_2022_NJP}, solid state systems~\cite{He_2017_Optica,Uppu_2020_Science,Knall_2022_PhysRevLett}, or superconducting circuits~\cite{Peng_2016_NatureComm,Zhou_2020_PhysRevApplied}. Photon-photon interactions can be achieved by harnessing intrinsic nonlinearities in these platforms, such as the Rydberg blockade in atomic ensembles~\cite{Lukin_2003_RevModPhys,Sangouard_2011_RevModPhys} or the anharmonic level structure of quantum dots~\cite{Gazzano_2016_JOptSocAmB}. However, such nonlinear interactions are typically weak at the single photon level. This challenge can be mitigated by integrating hybrid systems into cavity QED architectures, where the quantum emitter is strongly coupled to an optical cavity. The cavity enhances nonlinear interactions by strongly confining photons. Through the Purcell effect~\cite{Goy_1983_PhysRevLett}, it accelerates emission through a single cavity mode, improving both photon collection efficiency and indistinguishability, which are critical for interference-based quantum protocols~\cite{Hong_1987_PhysRevLett,Aaronson_2011_ACMSymp}. Conventional photonic sources based on spontaneous parametric down conversion (SPDC) and four wave mixing (SFWM)~\cite{Caspani_2017_Light} generate photons via inherently probabilistic processes, limiting their scalability for large scale quantum networks. In contrast, hybrid architectures offer a scalable pathway toward deterministic, on-demand photon generation, providing potential advantages for quantum communication and computing applications~\cite{Chang_2014_Nature,Slussarenko_2019_APR}.

Several experiments have demonstrated near deterministic single photon generation, including Ref.~\cite{magro_nature_photonics_2023}, which reports high efficiency creation of photonic states with non-classical features such as antibunching and Wigner function negativity~\cite{Curtright_2014}. This is achieved using a hybrid architecture consisting of an ensemble of Rubidium atoms coupled to a medium finesse optical cavity. The protocol leverages the Rydberg blockade mechanism~\cite{Jaksch_2000_PhysRevLett,Lukin_2001_PhysRevLett,Saffman_2010_RevModPhys} to induce strong optical nonlinearity~\cite{Firstenberg_2016_Journal_Phys_B}, enabling the coherent preparation of a collective Rydberg excitation delocalized across the ensemble. This collective excitation serves as a quantum memory, storing quantum information that is subsequently mapped onto a freely propagating photonic qubit. The system achieves a single photon emission efficiency of 60\%. The emitted photons inherit the intrinsic frequency stability and temporal control of the ensemble, while phase matching leads to enhanced, spatially selective emission into a well-defined cavity mode. Moreover, collective enhancement of light-matter interactions in atomic ensembles~\cite{Dudin_2012_Nature_Physics,Stanojevic_2009_PhysRevA} enables the use of moderate-finesse cavities, thereby relaxing the stringent high-finesse requirements that typically arise from the small interaction cross sections of single trapped atoms.

While this approach offers significant advantages, the system remains susceptible to decoherence mechanisms that fundamentally limit its performance. The two primary sources of decoherence during the excitation process are (i) the finite temperature of the ensemble, which induces random atomic motion and results in non-Markovian dephasing, and (ii) the imperfect Rydberg blockade, which permits unwanted double excitations within the ensemble~\cite{vaneecloo_phdthesis}. These effects degrade the coherence and single-excitation character of the collective Rydberg excitation, thereby disrupting the phase matching condition essential for photon emission into the desired cavity mode. To address these challenges, we employ quantum optimal control strategies to enhance the coherent preparation of the collective Rydberg excitations.

The exponential growth in computational complexity when modeling many-body quantum systems presents a significant challenge for optimizing the excitation process in such a large atomic ensemble. To address this, we begin by analyzing a reduced dimensional model based on collective Dicke states~\cite{Dicke_1954_PhysRev}, which captures the essential excitation dynamics without requiring simulation of the full $N$-atom density matrix. This approach models coherent dynamics using symmetric Dicke states and incoherent processes via asymmetric Dicke states, all within a Lindblad master equation framework~\cite{magro_nature_photonics_2023}. Notably, non-Markovian Gaussian dephasing is effectively mapped onto a Markovian description by including a minimal subset of asymmetric states in the system Hamiltonian, capturing the essential memory effects. The remaining states form an effective continuum, into which the system irreversibly decays through Markovian dephasing~\cite{Covolo_2025_Optica}. This formulation reflects an intrinsic source of decoherence that cannot be completely eliminated, though its impact can be partially mitigated through control. 

By monitoring the system dynamics, we identify double excitations at short timescales as the dominant decoherence mechanism. This insight motivates our use of a quantum control technique known as Derivative Removal for Adiabatic Gate (DRAG)~\cite{Motzoi_PhysRevLett_2009,Motzoi_PhysRevA_2013,Theis_2016_PhysRevA_1,Theis_2018_Europhysics_Letters}. The DRAG expansion algorithm, developed to suppress leakages, introduces analytical counterdiabatic corrections to the primary control field. These corrections reduce unwanted excitations and enable adiabatic state transfer at shorter timescales, improving the state transfer fidelity. It has been widely applied in superconducting platforms, effectively reducing microwave-induced leakages and crosstalk in transmon qubits~\cite{chow2010optimized,Chen_2016_PhysRevLett,Li_2024_npj,Li_2024_npj,li2024universal,wang2025suppressing, hyyppa2024reducing,jerger2024dispersive}. However, its application to atomic systems remains relatively underexplored~\cite{Theis_2016_PhysRevA_2,Petrosyan_2017_PhysRevA,singh2025optimizing}. In this work, we adapt the DRAG framework to shape optical pulses that drive excitations in a Rydberg ensemble, with the goal of suppressing unwanted double excitations and thereby mitigating one of the main sources of decoherence in the system.

Motivated by Ref.~\cite{magro_nature_photonics_2023}, we first determine the Hilbert space truncation required to capture the system dynamics, thereby identifying an optimal basis size for the Hamiltonian used in all the subsequent results. Using this converged model, we show that optimized DRAG pulses consistently enhance the single excitation probability compared to a simple sine-squared pulse by suppressing unwanted double excitations. Although the improvement is modest, it demonstrates that pulse shaping can partially mitigate leakage in the presence of strong Markovian dephasing. For context, we also analyze an idealized, dephasing-free limit. While this regime is not experimentally realistic, it provides a useful theoretical reference in which the same control strategy achieves substantially stronger suppression of double excitations. The contrast between this idealized case and the dephasing-dominated dynamics indicates that the limitations on attainable improvement arise primarily from Markovian dephasing rather than from the pulse optimization strategy itself. We further identify an optimal regime of pulse duration and ensemble size that enhances performance by reducing decoherence effects while remaining compatible with experimental constraints. Within this regime, the optimized DRAG pulses achieve a single excitation probability of $91.9\%$, surpassing the $77\%$ reported in Ref.~\cite{magro_nature_photonics_2023}. Importantly, this improvement is achieved without a significant reduction in ensemble size, thereby preserving the collective enhancements associated with large atom numbers. Finally, we benchmark DRAG against a fully numerical optimal control method, GRAPE~\cite{KHANEJA_2005}. GRAPE achieves comparable single excitation probabilities, indicating that DRAG already operates near the optimal control limit for the system while maintaining smooth, experimentally practical pulse shapes in contrast to the more complex GRAPE solutions. These results establish DRAG as an effective and experimentally feasible approach for improving single excitation preparation in Rydberg ensembles within the limits imposed by decoherence.

In Sec.~\ref{sec_model}, we introduce the system model, emphasizing the key decoherence mechanisms, the governing master equation, and the control objective. Section~\ref{sec_methods} details the DRAG optimization approach in both perturbative and non-perturbative forms, including the design of control pulses and the associated optimization methodology. In Sec.~\ref{sec_results}, we present and analyze the main results, focusing on the system’s response to DRAG, the optimization of pulse duration and atomic ensemble size, and benchmarking against GRAPE-based control. Finally, Sec.~\ref{sec_concl} summarizes the key findings and discusses their broader implications. Appendices provide supporting analyses, including the convergence analysis of the truncated Hilbert space required to capture the system dynamics (Appendix~\ref{app_convergence}), optimization in the idealized dephasing-free regime (Appendix~\ref{app_without dephasing}), value of the parameters used in the system model (Appendix~\ref{app_leakage}), and details on numerical simulation and optimization methods (Appendix~\ref{app_optimization}.


\section{A Rydberg Superatom in an optical cavity}
\label{sec_model}

\begin{figure}
    \centering
    \includegraphics[width=1\linewidth]{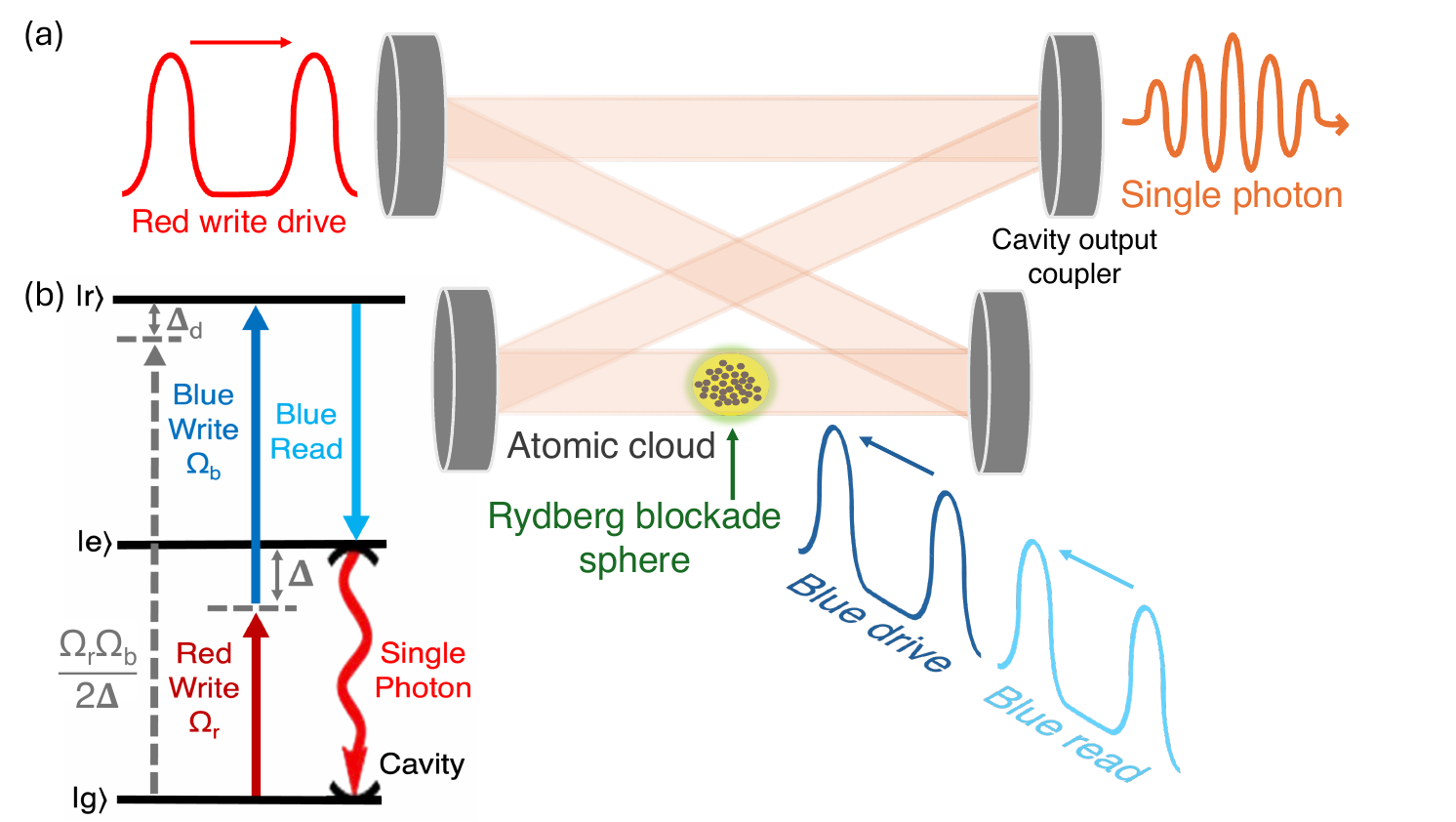}     
    \caption{\textbf{Schematic representation of the experimental setup.} (a) An atomic ensemble confined within an optical cavity, where the yellow shaded region denotes the Rydberg blockade sphere. (b) Two-photon excitation scheme in which a red drive (\(\Omega_r\)) and a blue drive (\(\Omega_b\)) coherently couple the ground state \( \ket{g} \) to the Rydberg state \( \ket{r} \) via a far-detuned intermediate state \( \ket{e} \). The gray dashed arrow indicates the effective two-photon Rabi coupling. Following excitation, a blue \textit{read} pulse drives a de-excitation process that results in the emission of a single photon into a well-defined cavity mode.}
    \label{fig:experimental_setup}
\end{figure}

The experiment described in Ref.~\cite{magro_nature_photonics_2023} features an atomic ensemble of approximately \( N \approx 800 \) cold \( ^{87}\mathrm{Rb} \) atoms, strongly coupled to a medium-finesse, single-ended optical cavity. A schematic of the setup is shown in Fig.~\ref{fig:experimental_setup}. Using well-established laser cooling and trapping techniques~\cite{Vaneecloo_2022_PhysRevX}, the atoms are cooled to a temperature of \( 5\,\mu\mathrm{K} \) and confined within a microscopic volume, as depicted in Fig.~\ref{fig:experimental_setup}(a). The ensemble has a Gaussian density profile with a root-mean-square (RMS) radius of \( \sigma = 4.6\,\mu\mathrm{m} \). Initially, all atoms are prepared in the ground state \( \ket{g} = \ket{5S_{1/2}, F=1, m_F=1} \). The excitation to the Rydberg state
\( \ket{r} = \ket{109S_{1/2}, J = 1/2, m_J = 1/2} \) is achieved via a two-photon transition driven by a pair of \textit{write} laser pulses. A red laser at 795~nm, with Rabi frequency \( \Omega_r \), and a blue laser at 475~nm, with Rabi frequency \( \Omega_b \), couple \( \ket{g} \) to \( \ket{r} \) via an intermediate excited state
\( \ket{e} = \ket{5P_{1/2}, F = 2, m_F = 2} \). Owing to the large detuning of the intermediate state, \( \Delta / (2\pi) = -500~\mathrm{MHz} \), the population in \( \ket{e} \) remains negligible. As a result, the dynamics can be effectively described by a two-photon Rabi frequency \( \Omega_r \Omega_b / (2\Delta) \); see Fig.~\ref{fig:experimental_setup}(b).

The strong interactions between the Rydberg atoms give rise to a blockade effect, wherein the energy levels of neighboring atoms are shifted out of resonance, thereby preventing multiple simultaneous excitations within the ensemble. As a result, the ensemble effectively behaves as a collective two-level system, often referred to as a \textit{superatom}, characterized by two fundamental collective states. The collective ground state is given by
\begin{equation} 
\ket{G} = \prod_{n=1}^{N_0} \ket{g_n},
\label{eq:Collective_G} 
\end{equation}
where all atoms remain in the ground state \( \ket{g} \). Here, \( \ket{g_n} \) indicates that the \( n \)-th atom is in the ground state, and \( N_0 = N / (1 + 4\sigma^2 / w^2) = 670 \) denotes the effective number of atoms that are optimally coupled to the cavity, where \( w = 21\,\mu\text{m} \) is the cavity waist. The collective singly-excited Rydberg state is given by
\begin{equation}
  \ket{R_0} = \frac{1}{\sqrt{N_0}} \sum_{j=1}^{N_0} e^{i \mathbf{k} \cdot \bm{r_j}}
  \left[ \prod_{n \neq j} \ket{g_n} \right] \ket{r_j},
\label{eq:Collective_R0}
\end{equation}
which describes a delocalized single excitation coherently shared across the entire atomic ensemble. This state corresponds to a symmetric superposition of all configurations in which a single atom is excited to the Rydberg state \( \ket{r} \) while all remaining atoms occupy the ground state \( \ket{g} \), and is invariant under the exchange of atoms. The collective nature of this excitation enhances the light-matter interaction, resulting in an effective Rabi frequency \( \Omega = \sqrt{N_0} \, (\Omega_r \Omega_b / 2 \Delta), \) which is increased by a factor of \( \sqrt{N_0} \) compared to the coupling strength in a single atom. The coherence of the state \( \ket{R_0} \) is encoded in the phase factor \( e^{i \mathbf{k} \cdot \bm{r_j}} \), where \( \bm{r_j} \) denotes the position of the \( j \)-th atom and \( \mathbf{k} = \mathbf{k}_r + \mathbf{k}_b \) is the effective wave vector of the two-photon excitation, with \( \mathbf{k}_r \) and \( \mathbf{k}_b \) representing the wave vectors of the red and blue laser fields, respectively. This collective phase coherence enforces momentum conservation through phase matching, leading to enhanced directional photon emission~\cite{Saffman_2002_PhysRevA}. Preserving the coherence of \( \ket{R_0} \) is therefore essential for efficient single photon generation in a desired cavity mode.

The \textit{write} beams drive coherent Rabi oscillations between \( \ket{G} \) and \( \ket{R_0} \), creating an atomic superposition state \( \cos(\theta/2)\ket{G} - \sin(\theta/2)\ket{R_0} \), where \( \theta \) denotes the qubit rotation angle. To convert this atomic excitation into a photon, a resonant blue \textit{read} laser couples \( \ket{r} \) to \( \ket{e} \) [see Fig.~\ref{fig:experimental_setup}(b)]. Owing to the strong coupling between the \( \ket{g} \leftrightarrow \ket{e} \) transition and the cavity mode, the atomic excitation is coherently mapped onto a single photon. The atom-cavity coupling strength is also enhanced by a factor of \( \sqrt{N_0} \)~\cite{bimbard_phdthesis}. The atomic superposition state is thus coherently mapped onto a photonic state, \( \cos(\theta/2)\ket{0} + \sin(\theta/2)\ket{1} \), yielding a fully controllable photonic qubit. The successful generation of such a qubit therefore critically depends on the high-fidelity preparation of the coherent atomic excitation. However, this excitation process is fundamentally limited by several decoherence mechanisms.

\begin{figure}
    \centering
    \includegraphics[scale=0.3]{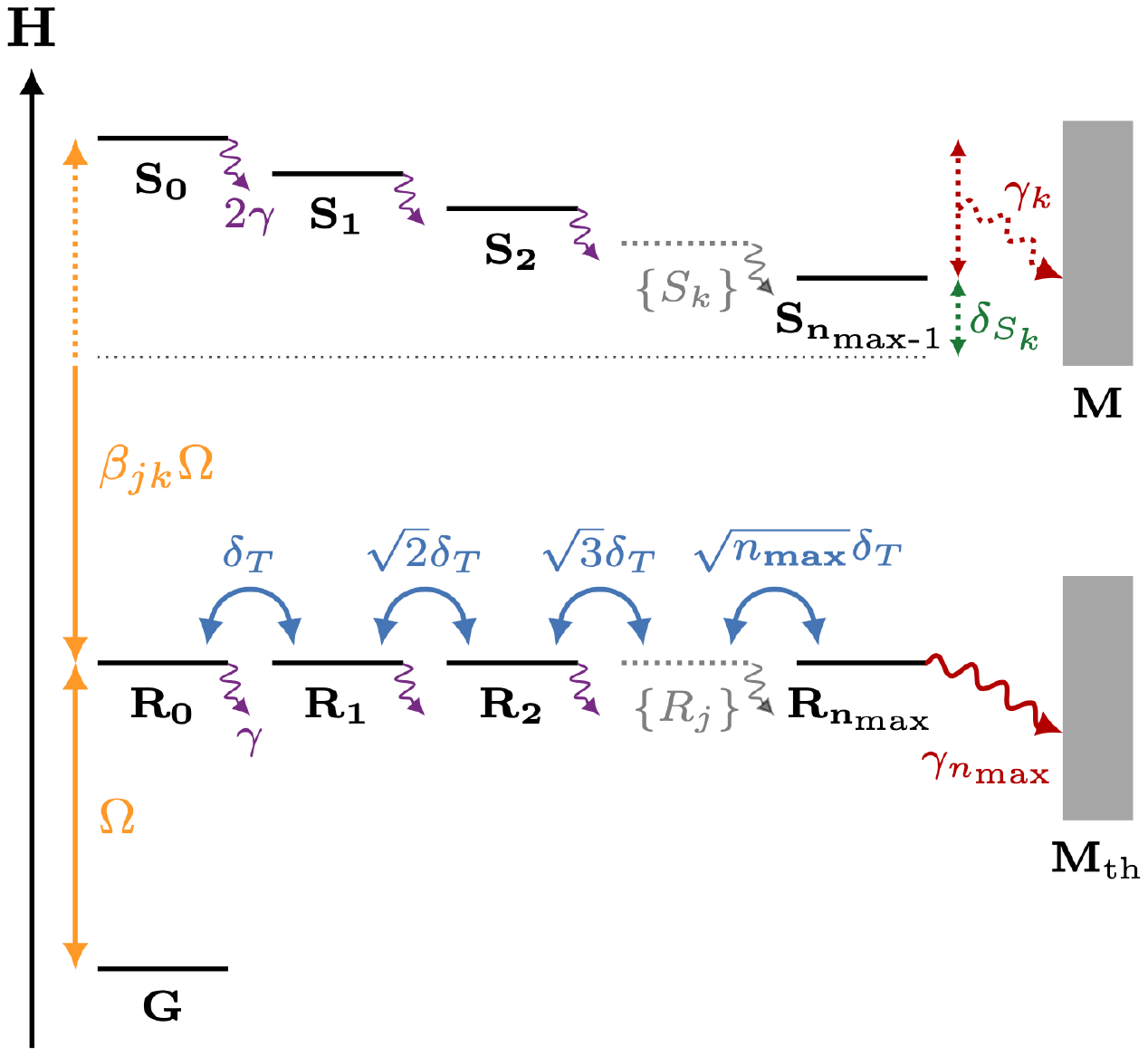}   
    \caption{\textbf{Schematic representation of the system dynamics.} The ground state $G$ couples to the singly-excited symmetric Rydberg state $R_0$ with Rabi frequency $\Omega$ (yellow arrow). The state $R_0$ sequentially couples to a series of singly-excited asymmetric Rydberg states $R_1, R_2, \!\ldots\!, R_j, \!\ldots\!, R_{n_{\text{max}}}$ (blue arrows), capturing memory effects arising from thermal atomic motion, and subsequently undergoes irreversible exponential decay (red arrow) into a continuum of these states $M_{\mathrm{th}}$ with a decay rate $\gamma_{n_{\text{max}}}$~\cite{Covolo_2025_Optica}. Each singly-excited asymmetric state further couples to the doubly-excited Rydberg states $S_0, S_1, \!\ldots\!, S_k, \!\ldots\!, S_{n_{\text{max}-1}}$ with corresponding Rabi frequencies $\beta_{jk}\Omega$ (yellow arrow). Each doubly-excited state is detuned by $\delta_{S_k}$ (green arrow) and irreversibly couples to the interaction-induced continuum $M$ with decay rate $\gamma_{S_k}$ (red arrow). All singly- and doubly-excited states share a common dephasing rate, $\gamma$ and $2\gamma$, respectively (purple arrows). This effective model explains the system dynamics while remaining computationally efficient.}
    \label{fig:fig_states}
\end{figure}

A significant source of decoherence in the system arises from the finite temperature of the atomic ensemble, which leads to random atomic motion. This motion induces a velocity-dependent phase shift on each component of $\ket{R_0}$ [Eq.~\eqref{eq:Collective_R0}], given by $e^{i \mathbf{k} \cdot \bm{v_j} t}$, where $\bm{v_j}$ is the velocity of the $j$-th atom. These velocity-dependent phases progressively destroy the phase-matching condition required for enhanced photon emission into a desired cavity mode. As a result, the collective excitation leaks out of the symmetric subspace into asymmetric singly-excited Rydberg states. Assuming that the atomic velocities follow a Maxwell-Boltzmann distribution at temperature $\mathcal{T}$, the resulting coherence loss of $\ket{R_0}$ follows a Gaussian profile~\cite{Zhao_2009_Nature}. The Doppler-induced frequency shifts responsible for this dephasing have a standard deviation $\delta_\mathcal{T} = |\mathbf{k}| \sqrt{k_B \mathcal{T} / m} = 2\pi \times 80 \, \text{kHz}$~\cite{magro_nature_photonics_2023}, where $k_B$ is the Boltzmann constant and $m$ is the atomic mass. This Gaussian dephasing is non-Markovian, meaning it retains memory of the atomic motion over time and cannot be accurately captured by a standard Lindblad master equation. However, as shown in Ref.~\cite{Covolo_2025_Optica}, such dephasing can be interpreted as a displacement in the time-frequency phase space and effectively incorporated into the Lindblad formalism by extending the Hilbert space to include a \emph{quasi-continuum} of asymmetric singly-excited Rydberg states. The symmetric state $\ket{R_0}$ couples coherently to a subset of these asymmetric states, $\{ \ket{R_j} \}$, with $j = 1, \dots, n_{\text{max}}$, capturing the essential memory effects with a coupling strength $\sqrt{j}\, \delta_\mathcal{T}$. It subsequently undergoes an irreversible exponential decay into a continuum of these states, denoted by $M_{\mathrm{th}}$, with a decay rate $\gamma_{n_{\text{max}}} = (2/\pi)^{(-1)^{n_{\text{max}}}/2} n_{\text{max}}!! / (n_{\text{max}}-1)!!$~\cite{Covolo_2025_Optica}, where $n_{\text{max}}$ is the index of the last asymmetric state included in the Hilbert space that couples directly to the continuum; see Fig.~\ref{fig:fig_states}. In essence, some states are treated with coherent evolution to capture memory effects, while the rest are treated using a dissipative Lindblad term to maintain computational efficiency.

The second major source of decoherence arises from imperfect Rydberg blockade, which occurs when the Rabi frequency exceeds the Rydberg-Rydberg interaction strength over a significant portion of the atomic ensemble. To model this effect, we adopt a lowest-order approximation in which at most two atoms are simultaneously excited. These atoms interact via a van der Waals potential, which shifts their resonance by \( \delta(r) = C_6 / r^6 \) as a function of their separation \( r \), where \( C_6 \) is the van der Waals coefficient characterizing the interaction strength. Since the atoms are randomly distributed, the number of doubly-excited states scales as \( \sim N_0^2/2 \), rendering a full numerical treatment computationally prohibitive. To make the problem tractable, we follow Ref.~\cite{magro_nature_photonics_2023} and model the atomic ensemble using a continuous Gaussian density profile. This approximation allows the spatial wavefunctions of the doubly-excited Rydberg states to be expanded in the basis of three-dimensional harmonic oscillator eigenstates. States for which the interaction potential diverges at short distances (due to the breakdown of the van der Waals approximation) are renormalized, and the interaction Hamiltonian is subsequently diagonalized within a subspace spanned by the lowest energy doubly-excited states. These states, denoted by \( \ket{S_k} \) with \( k = 0, \dots, n_{\text{max}} - 1 \), remain irreversibly coupled to a continuum of states outside the truncated subspace, labeled by \( M \); see Fig.~\ref{fig:fig_states}. Their coupling to this continuum gives rise to Lamb shifts and decay processes, from which we extract the interaction-induced energy shifts \( \delta_{S_k} \) and decay rates \( \gamma_{S_k} \). Finally, we compute the laser coupling coefficients \( \beta_{jk} \) between the singly-excited states \( \ket{R_j} \) and the doubly-excited states \( \ket{S_k} \), yielding a Rabi frequency for transitions between \( \ket{R_j} \) and \( \ket{S_k} \) of \( \beta_{jk} \Omega \). As in the case of thermal dephasing, this approach captures the essential physics of imperfect blockade by retaining some doubly-excited states in the system Hamiltonian, while treating the remainder as an effective continuum to maintain computationally efficiency. Additional sources of decoherence, including laser frequency noise, imperfect optical pumping, and residual uncompensated electric fields, are incorporated through an overall dephasing rate \( \gamma = 2\pi \times 40\,\text{kHz} \)~\cite{magro_nature_photonics_2023}.

We now introduce the master equation that governs the system dynamics. The state of the system is described by the density matrix \( \hat{\rho} \). We define the operators \( \hat{\sigma}_{ab} = \ket{a}\!\bra{b} \), which describe transitions between \( \ket{b} \) and \( \ket{a} \), and \( \hat{\sigma}_{aa} = \ket{a}\!\bra{a} \), which account for pure dephasing of \( \ket{a} \). The resulting equation takes the form

\begin{equation}
\begin{aligned}
    \mathcal{L}_w(\hat{\rho}) =\; & -i \left[ \frac{\hat{H}_w}{\hbar}, \hat{\rho} \right] + \sum_{k=0}^{n_{\text{max}}-1} 2\gamma_{S_k} \, \mathcal{D}[\hat{\sigma}_{M S_k}] \hat{\rho} \\
    & + 2\gamma_{n_{\text{max}}} \, \mathcal{D}[\hat{\sigma}_{M_{\mathrm{th}} R_{n_{\text{max}}}}] \hat{\rho} \\
    & + \sum_{j=0}^{n_{\text{max}}} 2\gamma \, \mathcal{D}[\hat{\sigma}_{R_j R_j}] \hat{\rho} + \sum_{k=0}^{n_{\text{max}}-1} 4\gamma \, \mathcal{D}[\hat{\sigma}_{S_k S_k}] \hat{\rho},
\end{aligned}
\label{eq:Master_eq}
\end{equation}

where the dissipator is defined as 
\[\mathcal{D}[\hat{\sigma}]\hat{\rho} = \hat{\sigma} \hat{\rho} \hat{\sigma}^\dagger - \frac{1}{2} \{\hat{\sigma}^\dagger \hat{\sigma}, \hat{\rho}\}.\]

The effective \textit{write} Hamiltonian is  
\begin{equation}
\begin{aligned}
    \frac{\hat{H}_w}{\hbar} = & \frac{\Omega}{2}\hat{K} + \frac{\Omega^*}{2}\hat{K}^\dagger - \sum_{k=0}^{n_{\text{max}}-1} \delta_{S_k} \hat{\sigma}_{S_k S_k} \\
    & -\delta_\mathcal{T} \sum_{j=1}^{n_{\text{max}}} \sqrt{j} (\hat{\sigma}_{R_{j-1}R_j} + \hat{\sigma}_{R_j R_{j-1}}),
\end{aligned}
\label{eq:effective_write_Hamiltonian}
\end{equation}

where $\Omega$ is the effective Rabi frequency and
\[\hat{K} = \hat{\sigma}_{G R_0} + \sum_{j=0}^{n_{\text{max}}} \sum_{k=0}^{n_{\text{max}}-1} \beta_{jk} \hat{\sigma}_{R_j S_k}.\]

Recall that \( n_{\text{max}} \) denotes the maximum number of states included in the Hilbert space, while the remaining states are treated as a continuum. For example, when \( n_{\text{max}} = 2 \), the Hilbert space contains two singly-excited asymmetric states, \( \ket{R_1} \) and \( \ket{R_2} \), which account for memory effects arising from atomic motion, and two doubly-excited states, \( \ket{S_0} \) and \( \ket{S_1} \), due to Rydberg-Rydberg interactions. The remaining states form the corresponding continua \( M_{\mathrm{th}} \) and \( M \), respectively. To illustrate the structure of the system dynamics and the associated control objective, we first present the effective \textit{write} Hamiltonian for \( n_{\text{max}} = 2 \); this minimal model serves as a pedagogical example. Capturing the full dynamics, however, requires a sufficiently large basis size. Accordingly, in Appendix~\ref{app_convergence} we perform a convergence analysis to identify the minimal Hilbert space truncation that accurately reproduces these effects. All numerical simulations and optimizations presented in this work are performed using the converged basis size. For \( n_{\text{max}} = 2 \), the Hamiltonian takes the form

\begin{align}
H_w=
\renewcommand{\arraystretch}{1.4}
\setlength{\arraycolsep}{1.3pt}
\left(
    \begin{array}{c|cccccc}
    & \textbf{G} & \textbf{R}_0 & \textbf{R}_1 & \textbf{R}_2 & \textbf{S}_0 & \textbf{S}_1 \\ \hline
    \textbf{G} & 0 & \frac{\Omega^*}{2} & 0 & 0 & 0 & 0 \\
    \textbf{R}_0 & \frac{\Omega}{2} & 0 & -\delta_\mathcal{T} & 0 & \frac{1}{2} \beta_{00} \Omega^* & \frac{1}{2} \beta_{01} \Omega^* \\
    \textbf{R}_1 & 0 & -\delta_\mathcal{T} & 0 & -\sqrt{2} \delta_\mathcal{T} & \frac{1}{2} \beta_{10} \Omega^* & \frac{1}{2} \beta_{11} \Omega^* \\
    \textbf{R}_2 & 0 & 0 & -\sqrt{2} \delta_\mathcal{T} & 0 & \frac{1}{2} \beta_{20} \Omega^* & \frac{1}{2} \beta_{21} \Omega^* \\
    \textbf{S}_0 & 0 & \frac{1}{2} \beta_{00} \Omega & \frac{1}{2} \beta_{10} \Omega & \frac{1}{2} \beta_{20} \Omega & -\delta_{S_0} & 0 \\
    \textbf{S}_1 & 0 & \frac{1}{2} \beta_{01} \Omega & \frac{1}{2} \beta_{11} \Omega & \frac{1}{2} \beta_{21} \Omega & 0 & -\delta_{S_1} 
    \end{array}
\right).
\label{eq:hamiltonian}
\end{align}

The overall control objective is the deterministic generation of single photons within a well-defined cavity mode, which requires optimizing the preparation of the collective singly-excited Rydberg state \( \ket{R_0} \). This entails maximizing the population in \( \ket{R_0} \) while minimizing undesired transitions to other states. Hereafter, the term \emph{population in a state} refers to the corresponding excitation probability. For example, the population in $\ket{R_0}$ denotes the single-excitation probability. Starting from the collective ground state \( \ket{G} \), the population is coherently driven to the symmetric singly-excited Rydberg state \( \ket{R_0} \) via a time-dependent control pulse \( \Omega(t) \). The states \( \ket{G} \) and \( \ket{R_0} \) thus define the computational subspace. However, \( \ket{R_0} \) is further coupled to the asymmetric singly-excited Rydberg states \( \ket{R_1} \) and \( \ket{R_2} \), which effectively capture the non-Markovian Gaussian dephasing induced by thermal atomic motion. Additionally, \( \ket{R_2} \) is further coupled to the thermal continuum \( M_{\mathrm{th}} \)~[see Eq.~\eqref{eq:Master_eq}]. It is important to note that these couplings are determined by fixed system parameters and thus cannot be modified through pulse shaping alone. All the singly-excited Rydberg states are further coupled to the doubly-excited Rydberg states \( \ket{S_0} \) and \( \ket{S_1} \) via the same control pulse \( \Omega(t) \), with each coupling scaled by the corresponding \( \beta \) factor. The dominant leakage pathway depends on both the coupling coefficients \( \beta \) and the interaction shifts \( \delta_{S_k} \). For \( n_{\text{max}} = 2 \), the dominant leakage channel is \( \ket{S_1} \), since \( \ket{S_0} \) is far off-resonant and its coupling coefficient is insufficient to induce significant population transfer. At short timescales (\( \Omega_{\mathrm{max}} \gg \delta_\mathcal{T} \)), the population in states \( \ket{R_1} \) and \( \ket{R_2} \) remain low; consequently, most of the leakage to \( \ket{S_1} \) originates from \( \ket{R_0} \). In addition, irreversible decay to the continuum \( M \) is also mediated via the doubly-excited states~[see Eq.~\eqref{eq:Master_eq}]. Therefore, minimizing population transfer from \( \ket{R_0} \) to \( \ket{S_1} \) is crucial for optimal system performance. Dephasing due to other common sources, characterized by \( \gamma \), is comparatively small. 

Although the discussion above refers to the \( n_{\text{max}} = 2 \) Hamiltonian [see Eq.~\eqref{eq:hamiltonian}], the same physical mechanism and control objective persist for larger Hilbert space truncations [see Fig.~\ref{fig:fig_states}]. In summary, the main sources of leakage are: (i) unitary transfer to the singly-excited asymmetric states \( \ket{R_1}, \dots, \ket{R_{ n_{\text{max}}}} \) at longer timescales due to atomic motion; (ii) unitary transfer to the leakage dominated doubly-excited state \(\ket{S_{\text{leak}}} \) at shorter timescales due to Rydberg-Rydberg interactions (\( \ket{S_{\text{leak}}} = \ket{S_1} \) for \( n_{\text{max}} = 2 \); for larger \( n_{\text{max}} \), a different doubly-excited leakage state dominates); (iii) non-unitary decay from \( \ket{R_{n_\text{max}}} \) to the thermal continuum \( M_{\mathrm{th}} \); and (iv) non-unitary decay from \( \ket{S_{\text{leak}}} \) to the interaction-induced continuum \( M \). To this end, we employ quantum control techniques to design a control pulse \( \Omega(t) \) that maximizes the state transfer fidelity to \( \ket{R_0} \) by reducing the dominant unitary leakage to the doubly-excited state \( \ket{S_{\text{leak}}} \), thus mitigating, to some extent, non-unitary leakage to \( M \). 

It is important to emphasize that all states introduced in the above modeling are auxiliary constructs; they do not correspond to distinct physical states in the experiment. Rather, they provide a conceptual and computationally efficient framework for a tractable description of decoherence in an open system.

\section{The Derivative Removal by Adiabatic Gate (DRAG) Method}
\label{sec_methods}

A broad range of quantum control techniques has been developed to address state transfer problems in complex quantum systems~\cite{Brif_2010_NJP, Glaser_2015_European_Physical_Journal_D, Koch_2022_EPJ}, spanning from gradient-based numerical optimization to model-free approaches such as machine learning. Among gradient-based methods, GRAPE~\cite{KHANEJA_2005} and its extensions~\cite{de2011second, motzoi2011optimal, goodwin2016modified, dalgaard2020hessian, dalgaard2022dynamical, jager2014optimal} are widely used to achieve high-fidelity control, while machine learning approaches have been explored primarily through reinforcement learning~\cite{dalgaard2020global, palittapongarnpim2017learning, an2019deep, nguyen2024reinforcement, Khalid2023} and meta learning techniques~\cite{preti2022continuous, chadwick2023efficient, kuzmanovic2025neural, de2024pulse, preti2022continuous}. In experimental settings, however, the performance of model-based optimization and offline training is often limited by noise, parameter uncertainties, and slow drifts. This has motivated the use of \emph{in-situ} calibration techniques based on direct experimental feedback. At the same time, the computational cost and noise sensitivity of high-dimensional optimization favor control strategies with only a few tunable parameters. Simple and flexible algorithms such as Nelder-Mead~\cite{nelder1965simplex} and related closed-loop methods like Chopped Random Basis (CRAB)~\cite{Caneva_2011_PhysRevA} are therefore commonly employed. Identifying pulse parametrizations that are both low-dimensional and sufficiently expressive remains challenging. Consequently, the most successful calibration strategies often combine few-parameter optimization with analytically motivated pulse constructions, enabling state-of-the-art gate fidelities across a range of platforms~\cite{Li_2024_npj,jerger2024dispersive, hyyppa2024reducing}.

One prominent example of an analytical pulse shaping technique is the Derivative Removal by Adiabatic Gate (DRAG) expansion, introduced in Refs.~\cite{Motzoi_PhysRevLett_2009, Motzoi_PhysRevA_2013, Theis_2018_Europhysics_Letters} to suppress leakage to non-computational states by incorporating a series of counterdiabatic corrections into the primary control field. This approach systematically diagonalizes unwanted Hamiltonian terms associated with distinct leakage pathways by applying a sequence of time-dependent unitary transformations. These transformations enable the construction of a counterdiabatic drive that cancels residual diabatic transitions, thereby allowing for fast and accurate quantum operations without relying on adiabatic evolution. Moreover, the method typically involves only a small number of tunable parameters, which can be experimentally calibrated to account for imperfections in the system model. A key advantage of this technique is that it yields continuous and relatively simple pulse shapes, making it well suited for direct experimental implementation.

In its original formulation~\cite{Motzoi_PhysRevLett_2009}, DRAG was developed to suppress diabatic leakage in a three-level system, where transitions between two computational states are accompanied by unwanted population transfer to a third, non-computational level. Notably, this suppression remained effective even in the presence of Markovian dephasing. A similar leakage mechanism exists in our system, where the desired excitation from the ground state \( \ket{G} \) to the singly-excited symmetric Rydberg state \( \ket{R_0} \) is accompanied by a dominant leakage into the doubly-excited Rydberg state \( \ket{S_{\text{leak}}} \). While our system includes additional states beyond the minimal three-level subspace, the dominant leakage pathway retains the same structure. Consequently, in this work, we adopt a similar class of parameterized DRAG pulse shapes to those presented in Refs.~\cite{Motzoi_PhysRevA_2013, li_PRXQuantum_2022}, as introduced below. 

For a single leakage channel, an off-quadrature auxiliary pulse, proportional to the scaled time derivative of the primary pulse, approximates the first order perturbative DRAG correction,
\begin{equation}
\Omega_{\text{DRAG}}^{\text{(pert)}}(t) = \Omega(t) - i\alpha \frac{\dot{\Omega}(t)}{\delta}.
\label{eq:Perturbative_drag}
\end{equation}
Here, \( \Omega_I = \Omega(t) \) is the primary in-phase component, and \( \Omega_Q = \alpha \dot{\Omega}(t) / \delta \) is the counterdiabatic out-of-phase component. \( \Omega_I (\text{and } \Omega_Q) \) induces rotations around the $x$-axis (and $y$-axis) in the computational subspace with an angle of rotation determined by the integral of the envelope \( \Omega_{\text{DRAG}}^{\text{(pert)}}(t) \). The parameter \( \alpha \) is the DRAG coefficient, which scales the amplitude of the counterdiabatic drive, and \( \delta \) is the energy gap between the target and leakage states.

While perturbative DRAG effectively suppresses leakage in the weak drive regime (\( \Omega < \delta \)), it becomes insufficient when \( \Omega \approx \delta \). This necessitates a non-perturbative extension to include higher-order corrections beyond Eq.~(\ref{eq:Perturbative_drag}). As detailed in Refs.~\cite{Li_2024_npj,li_PRXQuantum_2022}, non-perturbative DRAG employs the formalism of Givens rotations~\cite{Golub_1996} to iteratively diagonalize the Hamiltonian of the system, eliminating a single diabatic off-diagonal term at each step. These transformations enable the construction of a counterdiabatic drive that suppresses leakage beyond the limitations of perturbative corrections while ensuring convergence. The non-perturbative DRAG pulse for a single leakage channel is given by
\begin{equation} 
\Omega_{\text{DRAG}}^{\text{(non-pert)}}(t) = \Omega(t) - i\frac{\alpha}{\lambda} \frac{d}{dt} \tan^{-1} \left( \frac{\lambda \Omega(t)}{\delta} \right). 
\label{eq:Non_perturbative_drag} 
\end{equation}
Here, \( \lambda \) is a scaling factor for \( \Omega(t) \), characterizing the coupling strength to the leakage state.

In the laboratory setting, the physical DRAG control pulse applied to the system is given by
\begin{equation}
    \tilde{\Omega}(t) = \Omega_I(t) \cos(\omega_d t) + \Omega_Q(t) \sin(\omega_d t),
\end{equation}
where \( \omega_d \) is the common laser frequency that drives transitions in the system. For the considered Rydberg atom platform, these pulses can be experimentally generated by using a combination of an acousto-optic modulator (AOM) for amplitude modulation and an electro-optic modulator (EOM) for phase modulation.

Having established the DRAG framework, we now present the control pulse shapes and the optimization strategy used to maximize the population transfer to \( \ket{R_0} \). We begin with a primary in-phase control pulse, augment it with both perturbative and non-perturbative analytical DRAG corrections, and then further refine these pulses through numerical optimization.


We employ a sine-squared \(\pi\) pulse as the primary control pulse, defined as
\begin{equation}
    \Omega_{I}(t) = A \sin^2 \left( \frac{\pi t}{T} \right),
    \label{eq:Initial_pulse}
\end{equation}
where \( A \) is the amplitude and \( T \) is the pulse duration. The amplitude \( A \) is chosen such that the area of the pulse satisfies the \(\pi\) pulse condition, \( \int_{0}^{T} \Omega_I(t)\, dt = \pi, \) which yields \( A = 2\pi / T \). This pulse shape ensures that both \( \Omega_I(t) \) and its time derivative \( \dot{\Omega}_I(t) \) smoothly approach zero at the pulse boundaries \( t = 0 \) and \( t = T \), which is a prerequisite for implementing DRAG corrections. 


To maximize the population transfer to the target state \( \ket{R_0} \), we optimize the control pulse parameters by minimizing the loss function
\begin{equation}
    L = 1 - \left| \langle R_0 \mid \psi_f \rangle \right|^2 ,
    \label{Eq:Loss_function}
\end{equation}
where \( \ket{\psi_f} \) denotes the quantum state of the system at the final time. Starting from the initial state \( \ket{G} \), this loss function drives the dynamics toward the desired state \( \ket{R_0} \) while suppressing population in all the orthogonal states, in particular the dominant doubly-excited leakage state \( \ket{S_{\text{leak}}} \).


We begin by optimizing the pulse amplitude \( A \) and the laser detuning \( \Delta_d \), which represents a constant frequency offset between the drive field and the target transition (see Fig.~\ref{fig:experimental_setup}). Incorporating this detuning, the modified sine-squared control pulse takes the form
\begin{equation}
    \Omega_d(t) = \Omega_I(t)\, e^{i \Delta_d t},
    \label{eq:Initial_pulse_l}
\end{equation}
where the detuning induces a controlled phase evolution during the drive. This compensates for AC Stark shifts, allowing enhanced state transfer fidelity even prior to the application of DRAG corrections. 


To further suppress leakage into \( \ket{S_{\text{leak}}} \), we incorporate DRAG corrections into the primary control pulse using the analytical pulse profiles introduced in Eqs.~(\ref{eq:Perturbative_drag}) and (\ref{eq:Non_perturbative_drag}). As discussed earlier, the dominant leakage pathway in our system arises from the transition between \( \ket{R_0} \) and \( \ket{S_{\text{leak}}} \). Accordingly, the scaling factor \( \lambda \) is replaced by \( \beta_{0,\text{leak}} \), and the detuning parameter \( \delta \) by \( \delta_{S_{\text{leak}}} \). With these substitutions, the perturbative DRAG ansatz takes the form
\begin{equation}
\Omega_{\text{DRAG}}^{\text{(pert)}}(t)
= \left[ \Omega_I(t) - i \alpha \frac{\dot{\Omega}_I(t)}{\delta_{S_{\text{leak}}}} \right] e^{i \Delta_d t},
\label{eq:Perturbative_drag_1}
\end{equation}
while the non-perturbative DRAG ansatz is given by
\begin{equation}
\begin{split}
\Omega_{\text{DRAG}}^{\text{(non-pert)}}(t)
&= \Bigg[ \Omega_I(t) 
- \frac{i \alpha}{\beta_{0,\text{leak}}} \\
&\frac{d}{dt} \tan^{-1} \Bigg( 
\frac{\beta_{0,\text{leak}} \, \Omega_I(t)}{\delta_{S_{\text{leak}}}} 
\Bigg) \Bigg] e^{i \Delta_d t}.
\end{split}
\label{eq:Non_perturbative_drag_1}
\end{equation}
These physically motivated pulse shapes explicitly account for the dominant leakage dynamics and therefore provide well-informed initial ansätze for subsequent numerical optimization.


While the pulse shapes in Eqs.~(\ref{eq:Perturbative_drag}) and (\ref{eq:Non_perturbative_drag}) effectively suppress leakage to non-computational states, they do not fully eliminate phase errors within the computational subspace. Optimizing the constant laser detuning \( \Delta_d \) already reduces these errors substantially; however, additional improvement can be achieved by tuning the DRAG coefficient \( \alpha \). In standard DRAG implementations, \( \alpha \) is typically fixed to unity, a choice that minimizes leakage by canceling the spectral weight of the control pulse at the leakage transition frequency. This fixed value, however, does not by itself adequately compensate for Stark-induced phase errors. To address this limitation, we treat \( \alpha \) as an additional free parameter, alongside \( A \) and \( \Delta_d \), when optimizing the pulse shapes defined in Eqs.~(\ref{eq:Perturbative_drag_1}) and~(\ref{eq:Non_perturbative_drag_1}). This extended optimization strategy enables improved phase control within the computational subspace while preserving the suppression of unwanted transitions.


To avoid convergence to local minima, we employ a two-step numerical optimization procedure. A global search is first performed to broadly explore the control landscape and identify favorable regions in the parameter space, thereby providing robust initial conditions. The results of this search are then used to initialize a subsequent local optimization, which refines the pulse parameters and fine-tunes the control fields to maximize performance. The same loss function is used throughout to maintain consistency and enable direct comparison across all pulse shapes. Further details on the optimization methods are provided in Appendix~\ref{app_optimization}.

It is important to note that the effective control pulse \( \Omega(t) \), which drives transitions in the system and is being optimized here, is given by the product of the red and blue \textit{write} pulses, \(\Omega(t) = \sqrt{N_0} \Omega_r(t) \Omega_b / 2 \Delta\) [see Fig.~\ref{fig:experimental_setup}(b)]. In our optimizations, the blue \textit{write} pulse \( \Omega_b(t) \) is held constant, and only the red \textit{write} pulse \( \Omega_r(t) \) is varied. The amplitude and constant detuning are both controlled solely by adjusting the red \textit{write} pulse. Both \( \Omega_b \) and the prefactor \( \sqrt{N_0} / 2 \Delta \) remain constant throughout the optimization process and are effectively absorbed into the overall amplitude of the optimized control pulse.


\section{System Dynamics and Pulse Shaping}
\label{sec_results}

To systematically assess the performance of the DRAG control strategy, we first construct a reduced model that faithfully captures the essential system dynamics. As detailed in Appendix~\ref{app_convergence}, we perform a convergence analysis to determine the minimal Hilbert space truncation required, and find that a basis size of \( n_{\text{max}} = 8 \) is sufficient. This choice corresponds to explicitly including eight singly-excited asymmetric Rydberg states, \( \{\ket{R_1},\ldots,\ket{R_8}\} \), and eight doubly-excited Rydberg states, \( \{\ket{S_0},\ldots,\ket{S_7}\} \), in the system Hamiltonian. States beyond this truncation are treated as effective continua, \( M_{\mathrm{th}} \) and \( M \), which model irreversible thermal and interaction-induced dephasing, respectively. The resulting reduced Hilbert space therefore consists of 20 states, \( \{\ket{G}, \ket{R_0}, \ket{R_{1,\ldots,8}}, \ket{S_{0,\ldots,7}}, M_{\mathrm{th}}, M\} \), providing a computationally efficient yet faithful representation of the system dynamics. All parameter values required to simulate the Lindblad master equation~(\ref{eq:Master_eq}) for \( n_{\text{max}} = 8 \) are provided in Appendix~\ref{app_leakage}.

With this reduced model in place, we first investigate the system dynamics for the experimental pulse duration and ensemble radius reported in Ref.~\cite{magro_nature_photonics_2023}. Building on this analysis, we then identify an alternative parameter regime that exhibits enhanced resilience to decoherence and examine the corresponding dynamics. In both regimes, we assess and compare the performance of the four pulse shapes introduced earlier [see Eqs.~(\ref{eq:Initial_pulse}), (\ref{eq:Initial_pulse_l}), (\ref{eq:Perturbative_drag_1}), and (\ref{eq:Non_perturbative_drag_1})]. The pulse parameters are optimized following the strategy described in the preceding section. Finally, we perform GRAPE optimizations for both the experimental and optimized parameters as a benchmark. This comparison allows us to evaluate whether DRAG control already operates near the attainable performance limits or whether further improvements can be achieved using a fully numerical optimal control method.

\begin{figure*}[ht]
    \centering
    {\includegraphics[width=\textwidth]
    {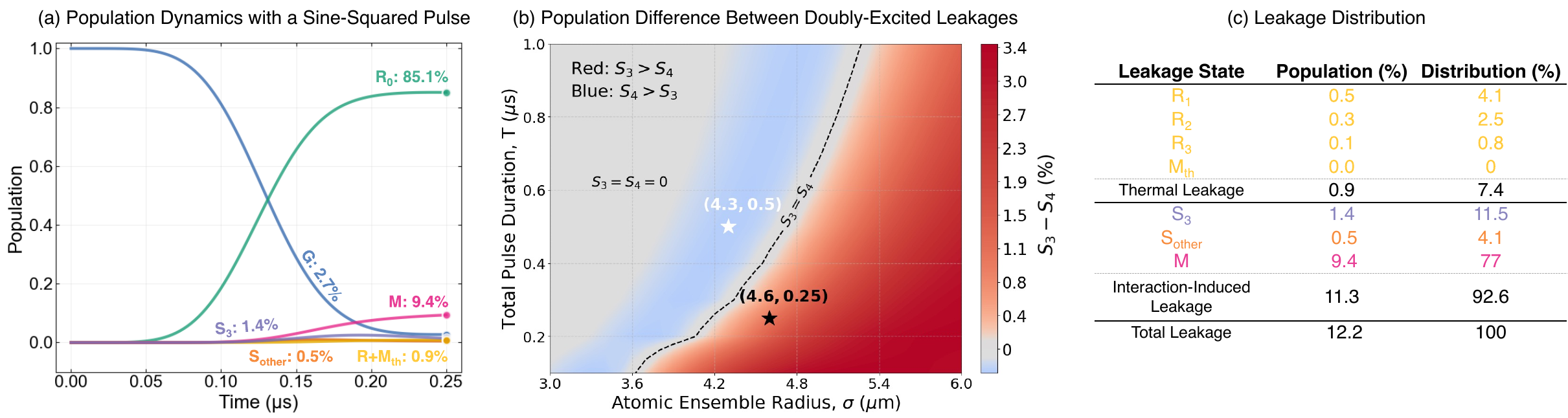}}
    \caption{\textbf{Identifying the dominant leakage channels.} (a) The system dynamics for \( T = 0.25~\mu\mathrm{s} \) and \( \sigma = 4.6~\mu\mathrm{m} \), corresponding to the experimental parameters of Ref.~\cite{magro_nature_photonics_2023} for a sine-squared $\pi$ pulse. Populations in the relevant states are shown: ground state \( G \) (blue), singly-excited symmetric state \( R_0 \) (green), all singly-excited asymmetric states and the associated thermal continuum \( R + M_{\mathrm{th}} \) (yellow), leakage dominated doubly-excited state \( S_3 \) (purple), remaining doubly-excited states \( S_{\mathrm{other}} \) (orange), and the associated interaction-induced continuum \( M \) (pink). Final populations are highlighted with matching colors. (b) Population difference between $S_3$ and $S_4$ as a function of the ensemble radius $\sigma$ and total pulse duration $T$, illustrating the dominant doubly-excited leakage state for different parameters. Colors indicate the sign of the difference;  $S_3>S_4$ in red and $S_4>S_3$ in blue. The black and white stars mark the experimental and optimized parameters, respectively. For the experimental parameters ($T=0.25~\mu\mathrm{s}$, $\sigma=4.6~\mu\mathrm{m}$), leakage is dominated by $\ket{S_3}$; for the optimized parameters ($T=0.5~\mu\mathrm{s}$, $\sigma=4.3~\mu\mathrm{m}$), it is dominated by $S_4$. (c) Final population distribution among the leakage states, normalized to the total leakage, for parameters in (a). The leakage into the doubly-excited state $S_3$ exceeds that into other doubly- and singly-excited asymmetric states, indicating $S_3$ as the dominant leakage channel and showing that interaction-induced dephasing dominates over thermal dephasing on short timescales. Population in $S_3$ irreversibly decays into the continuum $M$, and the difference between them quantifies the rapid Markovian dephasing.}
    \label{fig:leakage}
\end{figure*}

\subsection{System Dynamics under Experimental Parameters}  
\label{sec:expt_dynamics}

We begin by analyzing the system dynamics for the experimental setup corresponding to a pulse duration \( T = 0.25~\mu\mathrm{s} \) and an atomic ensemble radius of \( \sigma = 4.6~\mu\mathrm{m} \), see Fig.~\ref{fig:leakage}(a). As discussed in Sec.~\ref{sec_model}, the dominant doubly-excited leakage state is determined by the interplay between the coupling coefficients \( \beta_{jk} \), which govern the transitions between the singly-excited states \( \ket{R_j} \) and the doubly-excited states \( \ket{S_k} \), and the interaction-induced energy shifts \( \delta_{S_k} \). Since population transfer into the states \( \ket{S_k} \) occurs predominantly via the target state \( \ket{R_0} \), the couplings \( \beta_{0k} \) are of primary relevance. Both \( \beta_{0k} \) and \( \delta_{S_k} \) depend on the size of the truncated Hilbert space, characterized by \( n_{\text{max}} \). Throughout this work, we fix \( n_{\text{max}} = 8 \), which is sufficient to capture the relevant system dynamics [see Appendix~\ref{app_convergence}]. For this basis size, the interaction-induced energy shifts \( \delta_{S_k} \) decrease monotonically as one moves from \( \ket{S_0} \) to \( \ket{S_8} \) [see Eq.~(\ref{eq:energy_shifts}) in Appendix~\ref{app_leakage}]. At first glance, this behavior might suggest that the lowest energy state, \( \ket{S_8} \), should dominate the leakage dynamics due to its energetic proximity to \( \ket{R_0} \). However, the coupling coefficients \( \beta_{0k} \) also decrease with increasing \( k \) [see Eq.~(\ref{eq:coupling_coefficient}) in Appendix~\ref{app_leakage}], leading to a weaker effective coupling between \( \ket{R_0} \) and the lower-lying doubly-excited states. Consequently, although \( \ket{S_8} \) is energetically close to \( \ket{R_0} \), its coupling is strongly suppressed. Conversely, while \( \ket{S_0} \) is strongly coupled to \( \ket{R_0} \), it is significantly detuned. The dominant leakage channel therefore arises from a balance between detuning and coupling strength and typically corresponds to an intermediate energy state, such as \( \ket{S_3} \) or \( \ket{S_4} \). As Eq.~(\ref{eq:energy_shifts}) in Appendix~\ref{app_leakage} explicitly shows, the interaction-induced energy shifts \( \delta_{S_k} \) depend on the ensemble radius \( \sigma \); consequently, the dominant doubly-excited leakage state varies across the parameter space. For the considered experimental parameters, the doubly-excited state \( \ket{S_3} \) is the dominant leakage pathway. This behavior is illustrated in Fig.~\ref{fig:leakage}(b), which shows the population difference between the states \( \ket{S_3} \) and \( \ket{S_4} \) as a function of the atomic ensemble radius \( \sigma \) (in \( \mu\mathrm{m} \)) and the total pulse duration \( T \) (in \( \mu\mathrm{s} \)). For \( \sigma = 4.6~\mu\mathrm{m} \) and \( T = 0.25~\mu\mathrm{s} \) [indicated by the black star in the red region of Fig.~\ref{fig:leakage}(b)] the population difference is positive thus confirming \( \ket{S_3} \) as the dominant leakage channel. Accordingly, both perturbative and non-perturbative DRAG pulses are designed to suppress transitions into this channel by setting \( \beta_{0,\text{leak}} = \beta_{03} \) and \( \delta_{S_{\text{leak}}} = \delta_{S_3} \) in Eqs.~(\ref{eq:Perturbative_drag_1}) and~(\ref{eq:Non_perturbative_drag_1}).

As shown in Fig.~\ref{fig:leakage}(a), for a sine-squared pulse, the population transfer to the target state \( \ket{R_0} \) reaches \( 85.1\% \). This transfer is accompanied by a dominant leakage of \( 1.4\% \) into \( \ket{S_3} \), while the population in the remaining doubly-excited states \( \ket{S_{\mathrm{other}}} \) is negligible at \( 0.5\% \). The population in \( \ket{S_3} \) rapidly decays into the continuum \( M \) through a Markovian process, causing a substantial population loss of \( 9.4\% \). In addition, a smaller residual population of \( 0.9\% \) is transferred to the asymmetric singly-excited Rydberg states and the corresponding thermal continuum \( M_{\mathrm{th}} \) due to finite temperature effects in the atomic ensemble. This confirms the dominant leakage channel in the system. At shorter timescales, interaction-induced leakage is much larger than leakage due to thermal dephasing as seen from Fig.~\ref{fig:leakage}(c). Furthermore, the Markovian dephasing in the system is so rapid that the population entering \( \ket{S_3} \) is quickly transferred to \( M \), as evidenced by the pronounced population difference between \( \ket{S_3} \) and \( M \). 

\begin{figure*}
    \centering    
    {\includegraphics[width=\textwidth]{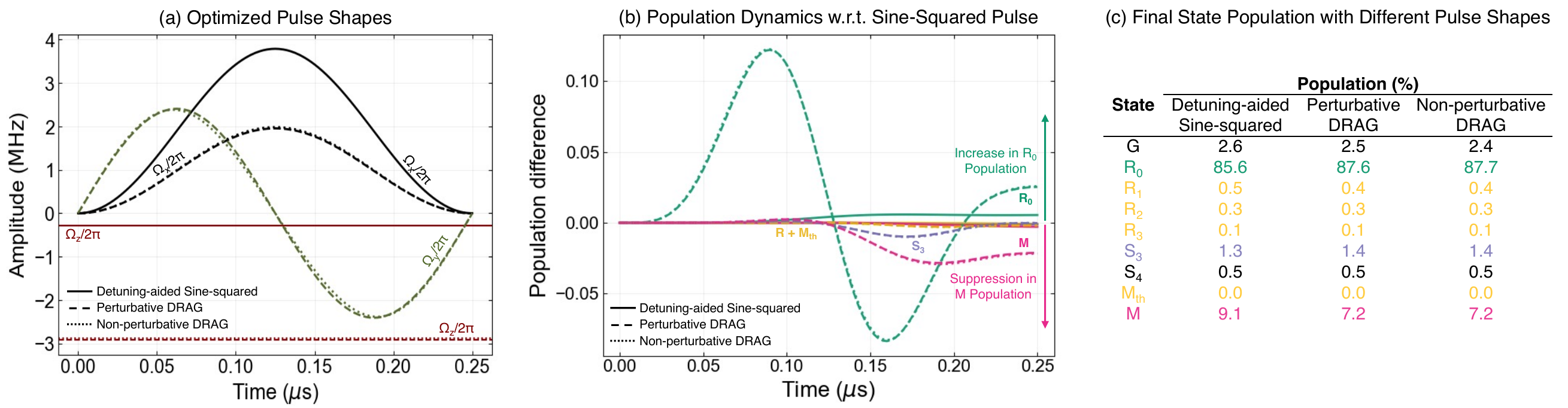}}    
   \caption{\textbf{System dynamics with the optimized pulses.} (a) Optimized control pulses for the detuning-aided sine-squared (solid), perturbative DRAG (dashed), and non-perturbative DRAG (dotted) schemes, showing the corresponding fields $\Omega_x$ (black), $\Omega_y$ (dark green), and $\Omega_z$ (dark red), which represent the in-phase (primary) drive, the out-of-phase (counterdiabatic) drive, and the constant laser detuning, respectively, for $T = 0.25~\mu\mathrm{s}$ and $\sigma = 4.6~\mu\mathrm{m}$. The optimized parameters for each pulse are: detuning-aided sine-squared: $A = 3.79$, $\Delta_d = -0.28$; perturbative DRAG: $A = 1.95$, $\Delta_d = -2.92$, $\alpha = -1.21$; and non-perturbative DRAG: $A = 2.00$, $\Delta_d = -2.87$, $\alpha = -1.20$. (b) Difference in the population dynamics relative to the initial sine-squared $\pi$ pulse for the optimized pulse shapes of panel (a) indicated with the same line code. Populations in the relevant states are shown; the singly-excited symmetric state $R_0$ (green), all singly-excited asymmetric states and the associated thermal continuum $R + M_{\mathrm{th}}$ (yellow), the leakage-dominated doubly excited state $S_3$ (purple), and the interaction-induced dephasing continuum $M$ (pink). Positive differences indicate the desired increase in the $R_0$ population with the optimized pulses, while negative differences indicate the targeted reduction in the population of $M$. The improvement provided by the DRAG pulses is higher than that obtained with the detuning-aided sine-squared pulse. (c) Final populations in all the relevant states for the optimized control pulses.}
   \label{fig:Initial_parameter_plots_dephasing_optimized} 
\end{figure*}

Introducing a detuning-aided sine-squared pulse [solid lines in Fig.~\ref{fig:Initial_parameter_plots_dephasing_optimized}(a)] yields modest improvements by increasing the population in the target state $\ket{R_0}$ to $85.6\%$ while reducing leakage into the continuum $M$ to $9.1\%$ [solid green and pink curves in Fig.~\ref{fig:Initial_parameter_plots_dephasing_optimized}(b)]. A higher suppression of leakage is achieved by incorporating DRAG corrections. Using perturbative DRAG [dashed lines in Fig.~\ref{fig:Initial_parameter_plots_dephasing_optimized}(a)], the $\ket{R_0}$ population increases to $87.6\%$, accompanied by a reduction of irreversible leakage into $M$ to $7.2\%$ [dashed green and pink curves in Fig.~\ref{fig:Initial_parameter_plots_dephasing_optimized}(b)]. The non-perturbative DRAG pulse [dotted lines in Fig.~\ref{fig:Initial_parameter_plots_dephasing_optimized}(a)] provides a slight additional improvement, reaching a final $\ket{R_0}$ population of $87.7\%$ [dotted green curve in Fig.~\ref{fig:Initial_parameter_plots_dephasing_optimized}(b)]. The populations in the remaining states for all the optimized pulses are summarized in Fig.~\ref{fig:Initial_parameter_plots_dephasing_optimized}(c).

Across all pulse shapes, populations in \( \ket{R_{1,\ldots,8}} + M_{\mathrm{th}} \) and the remaining doubly-excited states \( \ket{S_{\mathrm{other}}} \) remain small, validating the qualitative assumptions made in Sec.~\ref{sec_model}. In particular, the leakage to doubly-excited states occurs predominantly via $\ket{R_0}$, with $\ket{S_3}$ emerging as the dominant leakage channel and the primary source of irreversible population transfer to $M$. 

In addition, a residual population of approximately \( 2.5\% \) remains in the ground state \( \ket{G} \) for all the optimized pulse shapes. This residual population is not due to uncorrected phase errors but rather reflects an inherent trade-off in the optimization procedure. That is, further increasing the population transfer from \( \ket{G} \) to \( \ket{R_0} \) also enhances transitions to \( \ket{S_3} \), from where population rapidly decays into \( M \), precluding any possibility of recovery. To mitigate this effect, the optimization strategy favors a balanced approach that deliberately retains a small fraction of population in \( \ket{G} \) to suppress irreversible leakage and maximize the final population in \( \ket{R_0} \).

Overall, these results indicate that initializing the optimization with a DRAG-parametrized pulse is advantageous, resulting in modest but consistent improvements in the state transfer fidelity by suppressing spurious excitations into the dominant leakage channel and consequently reducing population in the continuum $M$. Compared to the standard sine-squared pulse, the DRAG pulse increases the final population in \( \ket{R_0} \) from 85.1\% to 87.7\%. The dephasing-free analysis presented in Appendix~\ref{app_without dephasing}, while experimentally unrealistic, shows that DRAG control performs near optimally in the absence of rapid Markovian dephasing. This comparison confirms that strong non-unitary dephasing represents the primary bottleneck for achieving higher population in \( \ket{R_0} \). In the presence of strong Markovian dephasing, the population accumulating in \( \ket{S_3} \) decays irreversibly into the continuum \( M \) before the corrective action of the DRAG pulse can take effect. In superconducting qubit platforms, the relevant dephasing rates are typically two to three orders of magnitude smaller than those considered in this model (see Appendix~\ref{app_leakage}), which explains why DRAG control is particularly effective in that setting. Motivated by these limitations, we next explore an alternative parameter regime defined by the atomic ensemble radius and total pulse duration) that offers enhanced robustness against decoherence.

\subsection{Optimization of the Parameters: Pulse Duration and Atomic Ensemble Radius}
\label{sec:parameter_opt}

\begin{figure*}
    \centering
    {\includegraphics[width=\textwidth, height=8.9cm]{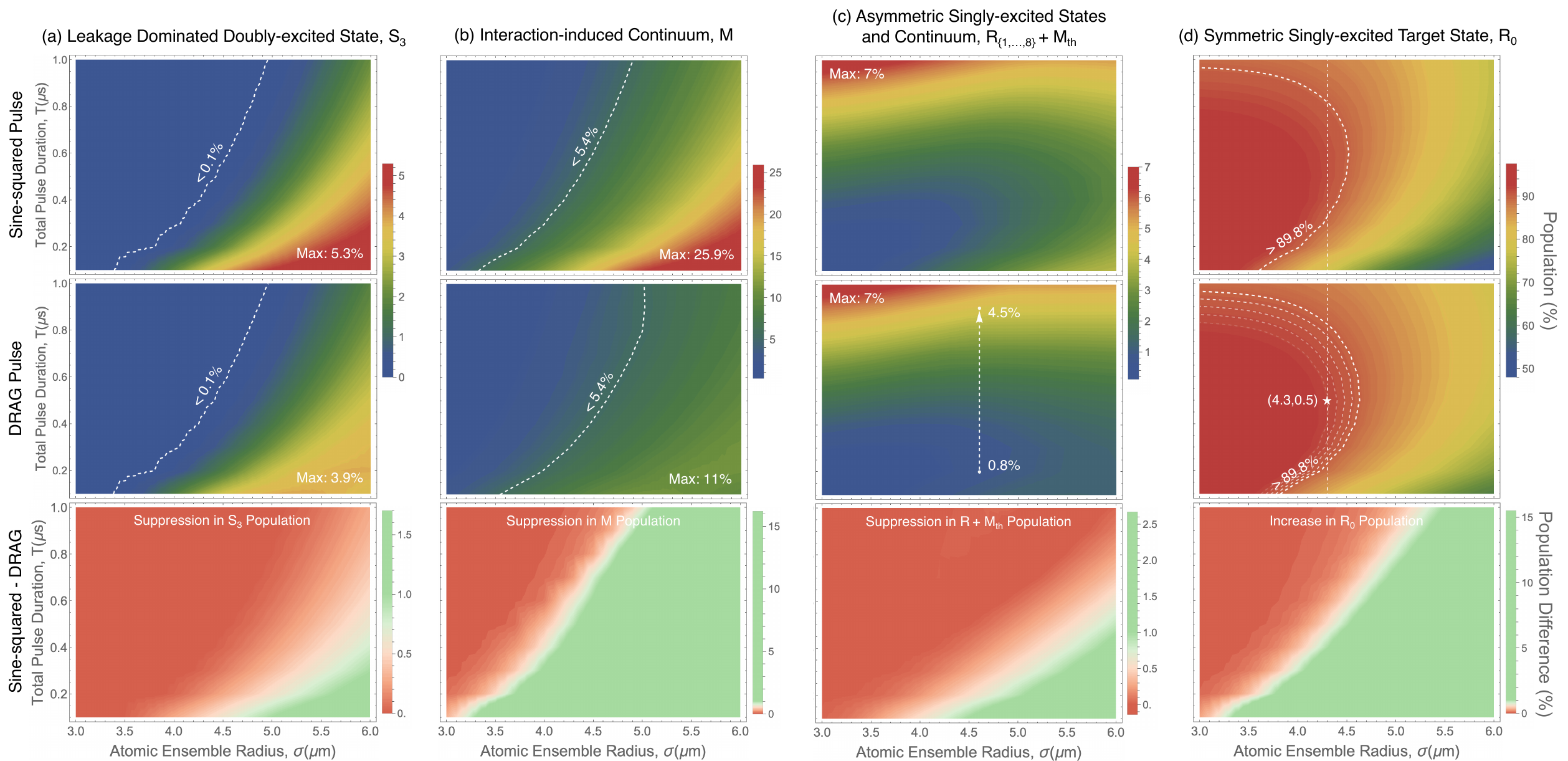}}
    \caption{\textbf{Identification of a parameter regime that minimizes dephasing.} Population distributions as functions of the atomic ensemble radius $\sigma$ and total pulse duration $T$, shown before DRAG optimization (first row) and after DRAG optimization (second row), together with their difference (third row), for (a) the leakage-dominated doubly excited state $\ket{S_3}$, (b) the interaction-induced dephasing continuum $M$, (c) the asymmetric singly-excited states and associated thermal continuum $\ket{R_{1,\ldots,8}} + M_{\mathrm{th}}$, and (d) the symmetric singly-excited target state $\ket{R_0}$. The color scale represents the final population after a complete pulse sequence, with red (dark blue) indicating high (low) population. These maps identify parameter regimes that minimize irreversible dephasing and highlight the effectiveness of DRAG in suppressing leakage. Leakage into $\ket{S_3}$ and $M$ increases with the ensemble size [see panels (a) and (b)] but is strongly suppressed by DRAG as indicated by the green regions in the third row, with optimal performance occurring to the left of the dashed contour curves. DRAG significantly enlarges the parameter space in which the leakage into $M$ remains below the $5.4\%$ threshold. In contrast, the population in $\ket{R_{1,\ldots,8}} + M_{\mathrm{th}}$ increases for longer pulse durations [see panel (c)] and is only weakly affected by DRAG. The population in the target state $\ket{R_0}$ is maximized for small ensemble radii and short pulses, a regime that is unfavorable due to collisional losses. The white star in the DRAG-optimized plot of panel (d) marks an optimal intermediate operating point that balances these competing constraints. For a lower bound of $\sigma = 4.3~\mu\mathrm{m}$, high-fidelity transfer to $\ket{R_0}$ is achieved at $T = 0.5~\mu\mathrm{s}$, as indicated by the dotted contour levels. For a threshold of $89.8\%$ population in $\ket{R_0}$, the accessible parameter space is again larger with DRAG.}
    \label{fig:contour_plots}   
\end{figure*}

To identify a parameter regime that minimizes the effects of decoherence, we analyze the final populations of four key states: the symmetric singly-excited target state $\ket{R_0}$, the asymmetric singly-excited states and the associated thermal continuum $\ket{R} + M_{\mathrm{th}}$, the leakage dominated doubly-excited state $\ket{S_3}$, and the associated interaction-induced dephasing continuum $M$. These populations are evaluated as functions of the total pulse duration $T$ and the atomic ensemble radius $\sigma$ for both the sine-squared and DRAG pulses, as shown in the first two rows of Fig.~\ref{fig:contour_plots}. The third row shows the population differences between the sine-squared and DRAG pulses, thereby highlighting the impact of DRAG on each state.

We begin by examining the behavior of the population in the state \( \ket{S_3} \) [see Fig.~\ref{fig:contour_plots}(a)], which increases with the radius of the atomic ensemble. This trend is attributed to the weakening of the Rydberg blockade in larger ensembles facilitating multiple excitations. DRAG pulse shaping effectively suppresses these unwanted excitations across the entire parameter space, particularly at larger radii, reducing the \( \ket{S_3} \) population by approximately 27\% in the bottom right corner of Fig.~\ref{fig:contour_plots}(a). A correlated trend is observed in the population in the interaction-induced dephasing continuum \( M \) [Fig.~\ref{fig:contour_plots}(b)], which is populated via transitions through \( \ket{S_3} \). As the population in \( \ket{S_3} \) increases, so does that of \( M \), contributing to coherence loss. The fact that the population in \( M \) is substantially larger than that in \( \ket{S_3} \) highlights the rapidity of the Markovian dephasing from \( \ket{S_3} \) to \( M \). The implementation of DRAG pulses significantly mitigates this effect, lowering the population in \( M \) by more than 57\% in the same region. Enhanced performance is achieved in the regions to the left of the dashed white contour curves in panels (a) and (b), where, after DRAG optimization, the populations in the leakage state \( \ket{S_3} \) and \( M \) is reduced to below 0.1\% and 5.4\%, respectively. Furthermore, for a fixed leakage threshold, the use of DRAG pulses significantly enlarges the accessible parameter space, as indicated by the rightward shift of the contour curve relative to that obtained with the sine-squared pulse in panel (b).

In contrast, the population in the states \( \ket{R_{1,\ldots,8}} \) and the thermal continuum \( M_{\mathrm{th}} \) [Fig.~\ref{fig:contour_plots}(c)] exhibits relatively weak sensitivity to DRAG optimizations. This behavior is expected because the transitions to these states are primarily driven by atomic motion at finite ensemble temperatures, a process that, as mentioned in Sec.~\ref{sec_model}, cannot be mitigated through pulse shaping alone. For longer pulse durations, the population in these states increases noticeably; it rises from approximately 0.8\% at \( T = 0.2~\mu\text{s} \) to about 4.5\% at \( T = 0.9~\mu\text{s} \) for \( \sigma = 4.6~\mu\text{m} \), as indicated by the dashed white arrow in the DRAG-optimized contour plot in Fig.~\ref{fig:contour_plots}(c). This trend reflects the increasing influence of atomic motion over extended timescales. Notably, DRAG pulses provide a reduction in the population leaking into these states at larger ensemble radii. This effect represents a secondary benefit. For a simple sine-squared pulse, the population initially leaks into the doubly-excited states from where it can subsequently redistribute into the asymmetric singly-excited states within the truncated Hilbert space. By suppressing the leakage into the doubly-excited manifold, DRAG pulses effectively cut off this feeder pathway resulting in an indirect reduction of population in these states. Importantly, DRAG does not directly counteract motional dephasing; rather, it prevents population from entering pathways that ultimately lead to population transfer into these states.

The target state \( \ket{R_0} \) exhibits maximal population in a regime characterized by short pulse durations and small atomic ensemble radii which benefits from both strong Rydberg blockade and reduced motional dephasing. For example, with a pulse duration of \( T = 100~\text{ns} \) and an ensemble radius of \( \sigma = 3~\mu\text{m} \), the population in \( \ket{R_0} \) reaches 97.4\% with a sine-squared pulse. 

Overall, we conclude that minimizing the population in \( \ket{S_3} \) and \( M \) favors smaller atomic ensemble radii due to the stronger Rydberg blockade. The achievable pulse duration is primarily constrained by the population in \( \ket{R_{1,\ldots,8}} \) and \( M_{\mathrm{th}} \), which decreases for shorter pulses as atomic motion is reduced. For example, if \( \ket{R_{1,\ldots,8}} \) and \( M_{\mathrm{th}} \) are neglected, longer pulse durations could be used to achieve higher population in \( \ket{R_0} \). In such an idealized scenario, weaker coupling to the doubly-excited states at longer \( T \) would allow the population in \( \ket{R_0} \) to reach 92.3\% for \( T = 1.2~\mu\mathrm{s} \) and \( \sigma = 4.6~\mu\mathrm{m} \) using a sine-squared pulse.

\begin{figure}
    \centering
    \includegraphics[scale=0.63]{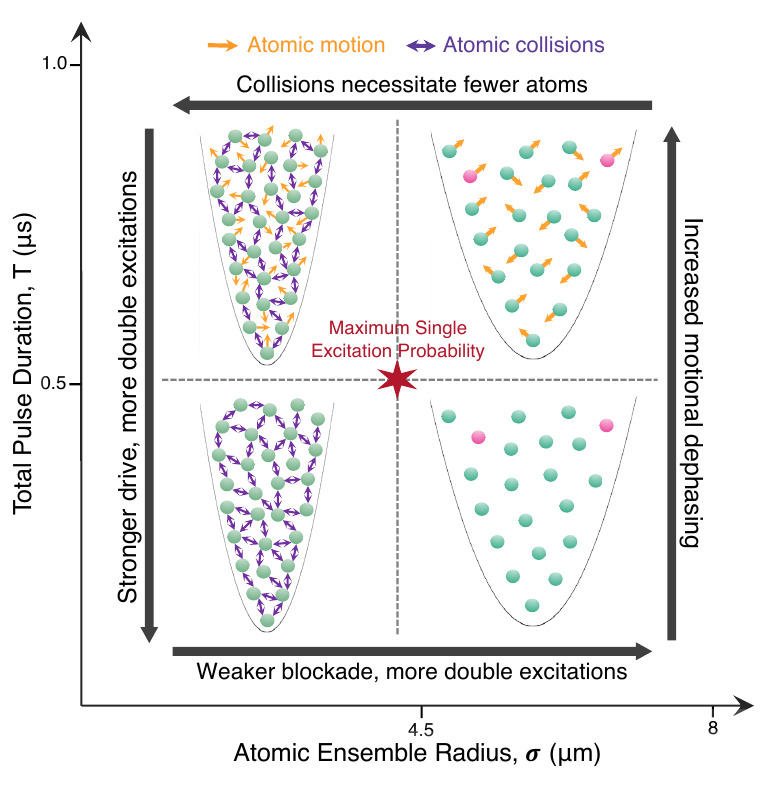}     
    \caption{\textbf{Different leakage mechanisms as a function of atomic ensemble radius and pulse duration.} An ensemble of atoms (green spheres) is confined in a harmonic trapping potential (solid black curve). Atomic motion and collisions are illustrated by yellow and purple arrows, respectively, while double excitations are indicated by pink spheres. As the ensemble radius \( \sigma \) increases, the Rydberg blockade weakens, leading to enhanced population in doubly-excited leakage states and, subsequently, in the continuum \( M \). Conversely, operation at small \( \sigma \) requires a reduction in the number of atoms in the ensemble to suppress collisional effects, which in turn reduces collective enhancements. Longer pulse durations allow additional time for atomic motion, thereby increasing population transfer to asymmetric singly-excited states and the associated thermal continuum \( M_{\mathrm{th}} \). In contrast, shorter pulses require stronger driving fields, which can inadvertently enhance coupling to the doubly-excited leakage states. The red star denotes an optimal operating point that balances these competing mechanisms, yielding maximal population transfer to the target state \( R_0 \).}
    \label{fig:parameter_trade-off}
\end{figure}

Taking all factors into account in the full model, the optimal regime appears to occur at small ensemble radii and short pulse durations. While this regime favors efficient excitation to \( \ket{R_0} \), additional practical limitations emerge. As the ensemble radius decreases, atomic density increases, enhancing the probability of collisions between Rydberg electrons and ground state atoms within the spatial extent of the Rydberg wavefunction~\cite{Bendkowsky_2009_Nature,Schlagmueller_2016_PRL}. These collisions introduce additional decoherence channels that are not accounted for in the current model. To mitigate such effects, the total number of atoms in the ensemble must be reduced at smaller radii. However, reducing the atom number diminishes collective enhancements, such as the enhanced Rabi frequency and atom-cavity coupling strength, which are crucial for efficient photon emission. Consequently, a careful balance must be maintained between maximizing the Rydberg blockade efficiency and preserving a sufficient atom number, avoiding excessive reduction of the ensemble size. The optimal performance is therefore achieved in an intermediate regime, where the atomic ensemble radius is neither too small, nor too large. Additionally, while shorter pulse durations are generally desirable to suppress motional dephasing, they become sub-optimal for atomic ensembles of moderate size. Achieving efficient excitation to \( \ket{R_0} \) with shorter pulses requires stronger driving fields, which, in turn, enhance unwanted coupling to the doubly-excited states, as illustrated by the white dashed contour curve in Fig.~\ref{fig:contour_plots}(d). To quantify these trade-offs [summarized in Fig.~\ref{fig:parameter_trade-off}], we analyze the population dynamics across the relevant parameter space. For pulse durations \( T < 0.4~\mu\mathrm{s} \), the required increase in the Rabi frequency leads to strong coupling to \( \ket{S_3} \), resulting in more than 4.9\% population leakage into \( \ket{S_3} \) and the continuum \( M \), even for a tightly confined ensemble with \( \sigma = 4.3~\mu\mathrm{m} \). Conversely, for longer pulses (\( T > 0.6~\mu\mathrm{s} \)), thermal atomic motion induces motional dephasing, causing population loss into \( \ket{R_{1,\ldots,8}} \) and, consequently, into \( M_{\mathrm{th}} \), exceeding 2\%. A similar constraint arises from the ensemble size: for radii \( \sigma > 4.6~\mu\mathrm{m} \), the weakened Rydberg blockade permits more than 5\% population in \( \ket{S_3} \) and \( M \), even for relatively long pulses (\( T = 0.6~\mu\mathrm{s} \)), where spectral broadening is reduced. Guided by this analysis, we identify moderate pulse durations (\( T = 0.4\text{-}0.6~\mu\mathrm{s} \)) and an ensemble radius of \( \sigma = 4.3~\mu\mathrm{m} \) as the optimal parameter regime, which enhances the Rydberg blockade while maintaining a sufficiently large atom number.

To illustrate the performance within this regime, we consider a representative case with \( T = 0.5~\mu\mathrm{s} \) and \( \sigma = 4.3~\mu\mathrm{m} \), corresponding to the white star in the DRAG-optimized contour plot of Fig.~\ref{fig:contour_plots}(d), where maximum population in \( \ket{R_0} \) can be achieved, as indicated by the dotted contour curves. For these parameters, the dominant doubly-excited leakage state is \( \ket{S_4} \), as shown in Fig.~\ref{fig:leakage}(b). The white star in the blue region of the figure marks the optimized parameters, where the population difference between \( \ket{S_3} \) and \( \ket{S_4} \) is negative. These parameters significantly reduce dephasing, even before the application of DRAG pulses. The population in $\ket{R_0}$ increases from 85.1\% with the experimental parameters to 91.4\% with the optimized parameters using a simple sine-squared pulse. Leakage to $\ket{S_4}$ and $\ket{M}$ is reduced to a combined 5.2\% with the optimized parameters, compared to 10.8\% with the experimental parameters. Non-perturbative DRAG pulse [with \( \beta_{0,\text{leak}} = \beta_{04} \) and \( \delta_{S_{\text{leak}}} = \delta_{S_4} \) in Eq.~(\ref{eq:Non_perturbative_drag_1})] further enhances population transfer, yielding a final population of 91.9\% in \( \ket{R_0} \)\footnote{For the optimized non-perturbative DRAG pulse, the final populations are $G = 2.4\%$, $R_0 = 91.9\%$, $R_1 = 1.3\%$, $R_2 = 0.2\%$, $S_3 = 0.1\%$, $S_4 = 0.3\%$, $M_{\mathrm{th}} = 0\%$, and $M = 3.8\%$.}. This represents one of the highest populations achieved using DRAG control within the practical constraints discussed above.

\subsection{Benchmarking against GRAPE}

\begin{figure*}
    \centering
    {\includegraphics[scale=0.44]
    {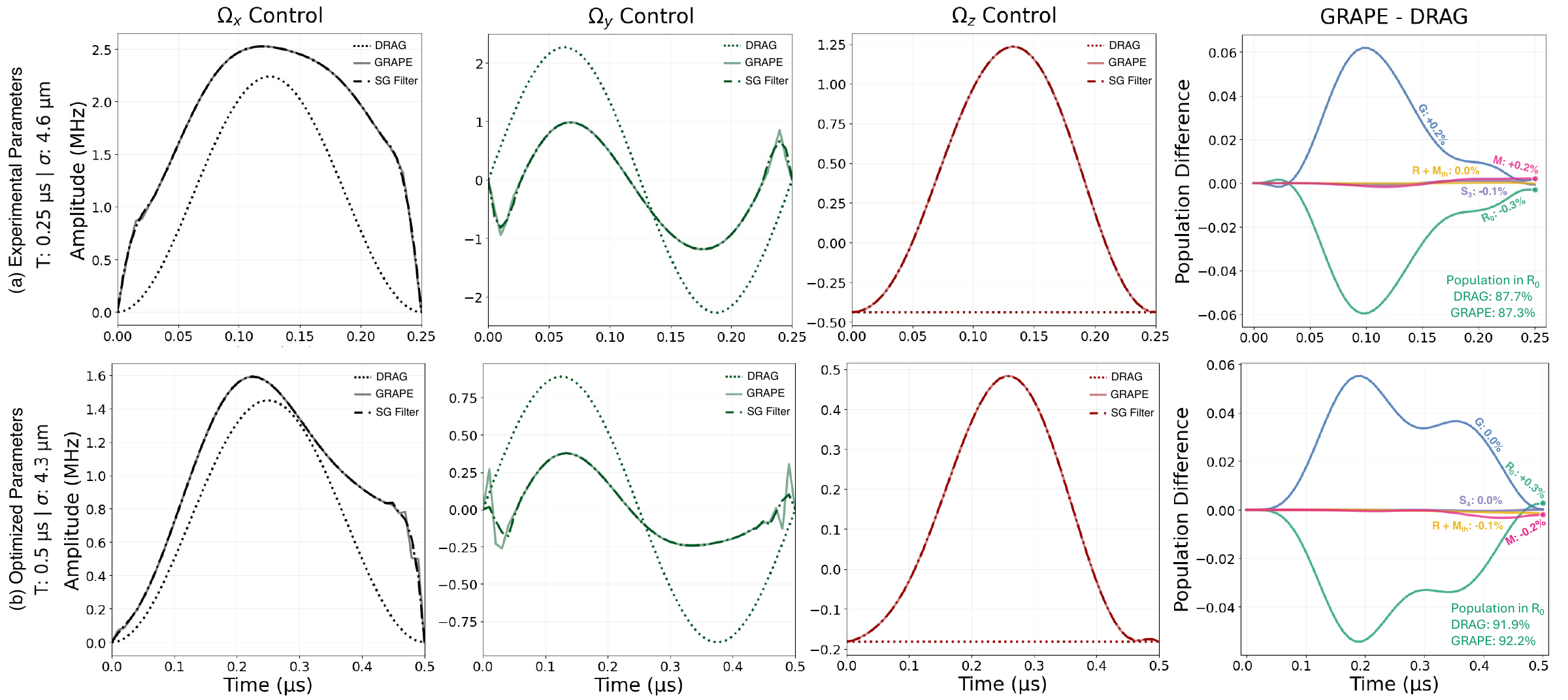}}
    \caption{\textbf{GRAPE-optimized control pulses and resulting system dynamics.} The first three panels show the control pulses $\Omega_x$ (black), $\Omega_y$ (dark green), and $\Omega_z$ (dark red), corresponding to the in-phase, out-of-phase, and detuning components. Each panel compares the initial DRAG (dotted), optimized GRAPE (solid), and SG-filtered GRAPE (dash-dotted) pulses. The fourth panel shows the difference in the dynamics generated by the SG-filtered GRAPE pulse and the DRAG pulse for (top) experimental and (bottom) optimized parameters. The shown populations include the ground state $G$ (blue), the singly-excited symmetric state $\ket{R_0}$ (green), all singly-excited asymmetric states and the associated thermal continuum $R + M_{\mathrm{th}}$ (yellow), the leakage dominated doubly-excited state $S_3$ or $S_4$ (purple), and the associated interaction-induced continuum $M$ (pink); final population differences are highlighted with matching colors. For the experimental parameters, GRAPE yields a final $\ket{R_0}$ population of $87.3\%$, which increases to $92.2\%$ for the optimized parameters. Comparison with DRAG control for both parameter sets shows minimal differences in the populations, indicating that DRAG already operates near the optimal control limit while retaining smooth and experimentally feasible pulse shapes.}
    \label{fig:GRAPE}
\end{figure*}

Next, we employ the GRAPE algorithm to optimize the in-phase (primary) and out-of-phase (counterdiabatic) components of the control field, together with the laser detuning. GRAPE discretizes the pulse and iteratively optimizes each time segment, using gradient information to efficiently find a high-fidelity solution in a high-dimensional parameter space. The optimization is initialized using the optimized DRAG pulse as the initial guess, which enables a direct assessment of the extent to which GRAPE can further improve upon the DRAG control. The control fields are discretized into 50 time steps, yielding a total of 150 optimization parameters corresponding to the in-phase, out-of-phase, and detuning components. To ensure experimental feasibility, we impose two constraints during the optimization. First, the pulse amplitudes are required to start and end at zero. Second, the change in amplitude between consecutive time steps is limited to less than \(0.5~\mathrm{MHz}\). Together, these constraints suppress strongly oscillatory solutions while retaining sufficient flexibility to achieve high-fidelity control. Under these conditions, the GRAPE-optimized pulses yield populations in the target state \(\ket{R_0}\) that are comparable to those obtained with the DRAG pulses. For the experimental parameters, corresponding to $T = 0.25~\mu\mathrm{s}$ and $\sigma = 4.6~\mu\mathrm{m}$, the final population in the target state $\ket{R_0}$ reaches $87.3\%$ using GRAPE, compared to $87.7\%$ using DRAG. For the optimized parameters, with $T = 0.5~\mu\mathrm{s}$ and $\sigma = 4.3~\mu\mathrm{m}$, the population increases to $92.2\%$ with GRAPE compared to $91.9\%$ with DRAG. Despite the regularization imposed by the constraints, the GRAPE-optimized pulses still exhibit small discontinuities arising from the time discretized nature of the optimization. To further suppress these features and obtain smoothly varying control fields, we apply the Savitzky-Golay (SG)~\cite{Savitzky_1964-co} filter to the optimized pulses. This post-processing step does not modify the final populations in \(\ket{R_0}\). Figure~\ref{fig:GRAPE} shows the control field components and the resulting differences in population dynamics between SG-filtered GRAPE and DRAG for both parameter sets. Notably, while the DRAG protocol employs a constant detuning and yields analytically defined smooth pulse shapes, the GRAPE optimization requires a fully time-dependent detuning. Despite this added flexibility, GRAPE does not provide a substantial improvement over DRAG. For comparison, we also perform unconstrained GRAPE optimizations initialized with random control pulses. While this approach is, in principle, well suited for estimating the theoretical upper bound on the achievable state transfer fidelity, it produces highly structured pulse shapes that are not experimentally feasible. Even in this case, the resulting fidelities remain comparable to those obtained using DRAG, with final populations of \(\mathrm{89\%}\) and \(92.2\%\) in \(\ket{R_0}\) for the experimental and optimized parameters, respectively. These results indicate that DRAG pulses already operate close to the optimal control limit set by decoherence while maintaining simple and experimentally feasible pulse forms.

\section{Summary and Conclusions}
\label{sec_concl}

In this work, we aimed to enhance the photon emission efficiency of a Rydberg-ensemble-based single-photon source by optimizing its excitation process through a detailed study of the model and identification of the main leakage channels. Efficient single photon generation requires high-fidelity preparation of a single Rydberg excitation while suppressing unwanted double excitations caused by imperfect Rydberg blockade. To address this limitation, we adapted the analytical DRAG pulse shaping technique to the Rydberg atom platform by numerically optimizing the control pulse parameters. While DRAG is shown to be effective at suppressing double excitations and provides modest improvements across the entire parameter regime, its performance is fundamentally limited by Markovian dephasing. Rapid population loss, driven by double excitations on short timescales and motional dephasing on longer timescales, emerges as the primary bottleneck restricting the overall system performance. Importantly, the longer timescale dephasing caused by thermal atomic motion is governed by parameters fixed in the system model and cannot be mitigated through pulse shaping, thereby constraining viable control pulses to relatively short durations. At the same time, excessively short pulses increase the likelihood of double excitations, as they require larger Rabi frequencies (\( \Omega \propto 1/T \)) and correspondingly higher laser power (\( \Omega \propto \sqrt{\text{Power}} \)) to maintain the same pulse energy. This broadens the spectral linewidth (\( \delta \nu \approx 1/T \)), reducing the effectiveness of the Rydberg blockade. An additional constraint arises from the spatial extent of the atomic ensemble. Reducing the ensemble radius strengthens the Rydberg blockade. However, this also increases the atomic density, raising the risk of collisional losses. Conversely, significantly reducing the atom number diminishes the benefits of collective enhancements, placing a lower bound on the optimal ensemble size. These competing factors define a narrow parameter regime in which high probability single excitation can be achieved. Within this optimal regime, we demonstrated a single excitation probability of 91.9\% using numerically optimized, non-perturbative DRAG pulse. This represents a substantial improvement over the 77\% theoretical probability obtained with the experimental parameters and pulse shapes reported in Ref.~\cite{magro_nature_photonics_2023}.

We also found that DRAG pulses operate near the optimal control limit, as evidenced by benchmarking them against pulses obtained via GRAPE. Despite its substantially greater flexibility, optimizing over 150 independent control parameters, GRAPE achieves performance comparable to that of DRAG. In contrast, DRAG pulses require optimization over only three parameters and are intrinsically smooth, making them both computationally efficient and experimentally convenient. To further probe the existence of a control limit, we tested the dressed Chopped Random Basis (dCRAB) method~\cite{Caneva_2011_PhysRevA}. In dCRAB, the control pulse is expressed as a truncated Fourier-basis expansion and optimized using a gradient-free direct search method, resulting in smooth pulse shapes while optimizing over 20 parameters (in this implementation), providing an intermediate level of flexibility between DRAG and GRAPE. Using dCRAB with the experimental parameters \( T = 0.25~\mu\mathrm{s} \) and \( \sigma = 4.6~\mu\mathrm{m} \), we achieved a single excitation probability of \( 87.6\% \). In the ideal dephasing-free limit [see Appendix~\ref{app_without dephasing}], dCRAB yielded lower fidelity than DRAG, underscoring the efficiency of few-parameter analytical solutions. Meanwhile, for the optimized parameters \( T = 0.5~\mu\mathrm{s} \) and \( \sigma = 4.3~\mu\mathrm{m} \), the single excitation probability reached \( 91.9\% \)~\cite{Thomas_QuOCS_2022}. These results further confirm that the achievable state-transfer fidelity is primarily limited by the system's intrinsically fast decoherence. Previous works applying DRAG to atomic systems~\cite{Theis_2016_PhysRevA_2,Petrosyan_2017_PhysRevA} report dephasing rates proportional to $1/\Gamma \approx 10^{-9} n^3$~s, roughly two to three orders of magnitude smaller than the dephasing rates considered in this model [see appendix~\ref{app_leakage}]. Nevertheless, DRAG still provides modest improvements even in the presence of such rapid decoherence.

To further enhance the single excitation probability beyond pulse shaping, we consider the influence of other experimentally adjustable parameters. One approach is to reduce the ensemble temperature, which directly suppresses motional dephasing. However, the achievable cooling is fundamentally limited by heating from the trapping and excitation beams, particularly during extended experimental durations. Moreover, excessive cooling would degrade the experimental duty cycle. An alternative strategy is to increase the van der Waals interaction strength by raising the principal quantum number \( n \) of the Rydberg level, since \( C_6 \propto n^{11} \). Stronger interactions enhance the Rydberg blockade and suppress double excitations. For example, at \( n = 200 \), where \( C_6 = 2\pi \times 155{,}563 \, \text{THz} \, \mu\text{m}^6 \)~\cite{SIBALIC_ARC_2017}, a simple sine-squared \( \pi \)-pulse of duration \( T = 0.25 \, \mu\text{s} \) applied to an ensemble of size \( \sigma = 4.6 \, \mu\text{m} \) (experimental parameters) yields an estimated 98.6\% population in the target state \( \ket{R_0} \). However, such high-lying Rydberg states present significant drawbacks, including increased sensitivity to stray electric fields (\( \propto n^7 \)), enhanced blackbody-induced transitions (\( \propto n^2 \)), and stronger level mixing due to Stark (\( \propto n^4 \)) and Zeeman (\( \propto n^2 \)) effects~\cite{Saffman_2010_RevModPhys}, which limit their practical viability. Additionally, the energy spacing between nearby Rydberg levels decreases with increasing \( n \), although this can be mitigated using DRAG~\cite{Theis_2016_PhysRevA_2}. A more balanced choice, such as \( n = 120 \) with \( C_6 = 2\pi \times 462 \, \text{THz} \, \mu\text{m}^6 \), combined with a pulse duration of \( T = 0.5 \, \mu\text{s} \) and a ensemble radius of \( \sigma = 4.6 \, \mu\text{m} \), yields over 95.40\% population in \( \ket{R_0} \) using simple pulse shapes. This choice enables high-fidelity single excitation while avoiding many of the challenges associated with extremely high Rydberg levels.

Finally, to address the origin of double excitations, we note that weakly interacting atoms that limit the effectiveness of the Rydberg blockade are predominantly located in the tails of the ensemble’s Gaussian density distribution. Experimentally removing these atoms could therefore enhance the blockade. Although this would slightly reduce the collective enhancement due to the smaller atom number, the resulting improvement in the blockade efficiency is expected to outweigh this reduction. In parallel, certain aspects of the model describing imperfections of the Rydberg blockade in a Gaussian ensemble have been refined. While these refinements lead to improved agreement with experimental data~\cite{Magro2026}, the overall structure of the model remains unchanged and the modifications (which are currently under investigation) are not expected to significantly affect the conclusions presented here. Together, these results highlight the delicate balance of competing factors governing high-fidelity single excitation generation in Rydberg ensembles and provide a roadmap for further improvements in single photon sources based on strongly interacting Rydberg atoms.

\begin{acknowledgments}

We acknowledge funding from the Horizon Europe programme HORIZON-CL4-2022-QUANTUM-02-SGA via the project 101113690 PASQuanS 2.1, from the programme “Profilbildung 2022” of the Ministry of Culture and Science of the State of North Rhine-Westphalia via the project “Quantum-based Energy Grids (QuGrids)”, as well as by Germany’s Excellence Strategy – Cluster of Excellence Matter and Light for Quantum Computing (ML4Q2) EXC 2004/2 – 390534769. E.C. was also supported by JSPS KAKENHI grant number JP23K13035. A.O. acknowledges the Plan France 2030 project ANR-22-PETQ-0013, and the CIFAR Azrieli Global Scholars program.

\end{acknowledgments}

\appendix

\appendix
\renewcommand{\thefigure}{\thesection.\arabic{figure}}

\section{Convergence Analysis of the Truncated Hilbert Space}
\label{app_convergence}
\setcounter{figure}{0}

\begin{figure}
    \centering
    \includegraphics[scale=0.55]{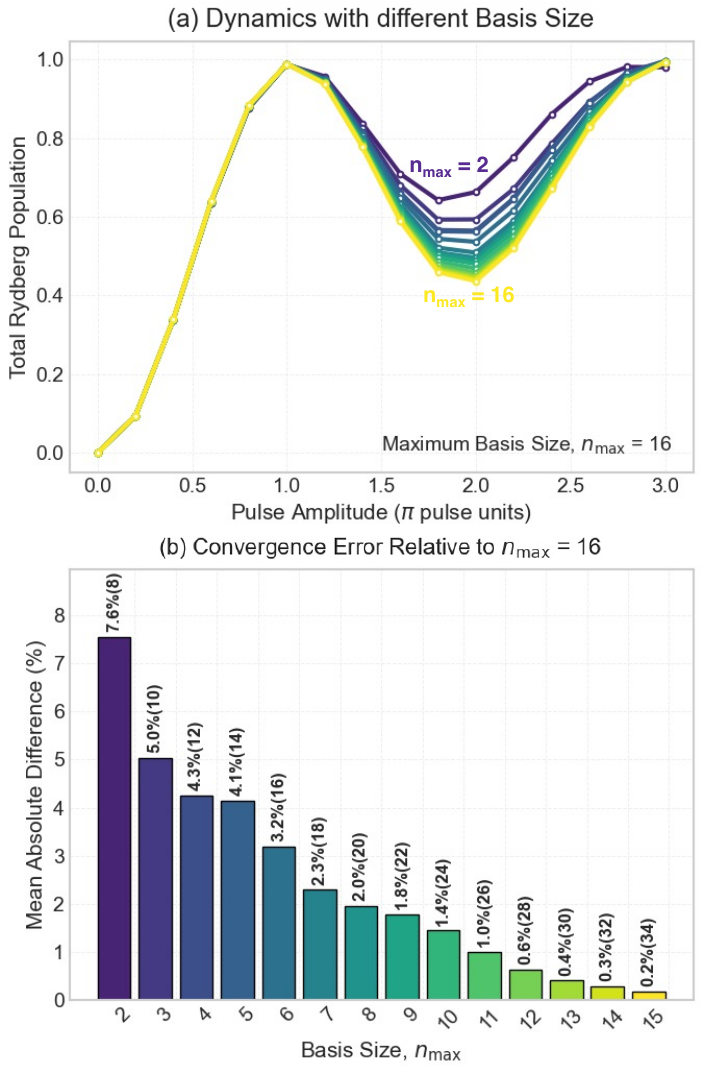}
    \caption{\textbf{Convergence analysis of the truncated Hilbert space.} (a) Total Rydberg population as a function of drive strength for different basis sizes, obtained by increasing the amplitude of a sine-squared $\pi$ pulse. The curves correspond to basis sizes ranging from $n_{\max} = 2$ (purple) to $n_{\max} = 16$ (yellow). Larger basis sizes reduce artificial damping of Rabi oscillations, providing a more faithful representation of the system dynamics. (b) Convergence of the total Rydberg population with respect to the basis size, quantified by the mean difference relative to the largest basis ($n_{\max} = 16$) across all Rabi frequencies. The difference with respect to $n_{\max}=16$ is shown above each bar plot, with the corresponding Hamiltonian size given in parentheses. A basis size of $n_{\max} = 8$ provides a practical compromise, capturing the essential dynamics while remaining computationally efficient.}
    \label{fig:convergence}
\end{figure}

As discussed in Sec.~\ref{sec_model}, the system dynamics involve population transfer from the symmetric subspace, spanned by $\ket{G}$ and $\ket{R_0}$, into the asymmetric subspace in order to capture non-Markovian thermal dephasing. Accurately modeling these dynamics requires retaining a sufficient number of asymmetric states in the Hilbert space, while treating higher-lying states as an effective continuum. This strategy enables the simulation to capture the essential memory effects while remaining computationally efficient. The same approach applies to doubly-excited states: a subset of these states is included explicitly in the system Hamiltonian, while the remainder is treated as a continuum, making the modeling of interaction-induced dephasing computationally efficient. To examine the dependence of the dynamics on the basis size \( n_{\max} \), we plot the total Rydberg excitation probability (i.e., $1 -$ population in \( \ket{G} \)) as a function of increasing drive strength for different basis sizes. Here, the basis size refers to the number of asymmetric singly- and doubly-excited Rydberg states explicitly included in the system Hamiltonian; for example, \( n_{\max}=2 \) corresponds to two singly-excited asymmetric states and two doubly-excited states, and so on. This yields a total of $2n_{\max}+4$ states in the system Hamiltonian, namely
\(
\left\{
\ket{G},\,
\ket{R_0},\,
\ket{R_{1,\ldots,n_{\max}}},\,
\ket{S_{0,\ldots,n_{\max}-1}},\,
M_{\mathrm{th}},\,
M
\right\}
\). As the drive strength increases, the Rydberg blockade is weakened, enabling double excitations and allowing the population to leak from the symmetric subspace into the asymmetric manifold. In a truncated Hilbert space, the population entering the asymmetric subspace can irreversibly decay into the continuum, thereby preventing recovery to the symmetric subspace. This mechanism manifests as damping of the Rabi oscillations between \( \ket{G} \) and \( \ket{R_0} \), as is clearly visible in Fig.~\ref{fig:convergence}(a). After a single \( \pi \)-pulse, the damping is strongest for the smallest basis size (\( n_{\max}=2 \)) and weakest for the largest one (\( n_{\max}=16 \)). For \( n_{\max}=16 \), most of the population coherently returns to the symmetric subspace, as indicated by the pronounced decrease in the total Rydberg population, whereas for \( n_{\max}=2 \) the recovered population is substantially reduced. These results demonstrate that increasing the basis size captures more faithfully the system dynamics and suppresses the artificial damping introduced by excessive Hilbert space truncation. All curves eventually return to unity at \( \Omega = 3\pi \), reflecting the total population conservation across the symmetric and asymmetric singly- and doubly-excited states, as well as the continua.

To determine the minimal basis size that captures these dynamics, we analyze the convergence of the total Rydberg population as a function of the basis size. For each truncated basis, we compute the difference in the total Rydberg population relative to that obtained with the largest basis used in this work (\(n_{\max}=16\)) and take the mean of this difference across all Rabi frequencies. Figure~\ref{fig:convergence}(b) shows these mean differences as a function of \(n_{\max}\). The mean difference decreases monotonically with increasing basis size, indicating that the artificial damping introduced by smaller Hilbert spaces diminishes as the Hilbert space grows. We choose \(n_{\max}=16\) as the maximal basis size for the convergence analysis, as it is sufficiently large to capture the essential dynamics along the asymmetric ladder while remaining computationally tractable. For subsequent pulse optimizations, we employ a reduced basis size of \(n_{\max}=8\) to lower the computational cost of repeated simulations, particularly for generating the contour plot in Fig.~\ref{fig:contour_plots}. Although the total Rydberg population at this truncation differs from that at \(n_{\max}=16\) (mean difference \(\sim 2\%\)), the key features of the population dynamics observed experimentally are well reproduced. We therefore adopt \(n_{\max}=8\) as a practical compromise that captures the relevant physics while enabling efficient optimization of control pulses, balancing simulation accuracy with computational resource requirements.

\section{Dephasing-free Regime}
\label{app_without dephasing}
\setcounter{figure}{0}

\begin{figure}
    \centering
    {\includegraphics[width=\linewidth]
    {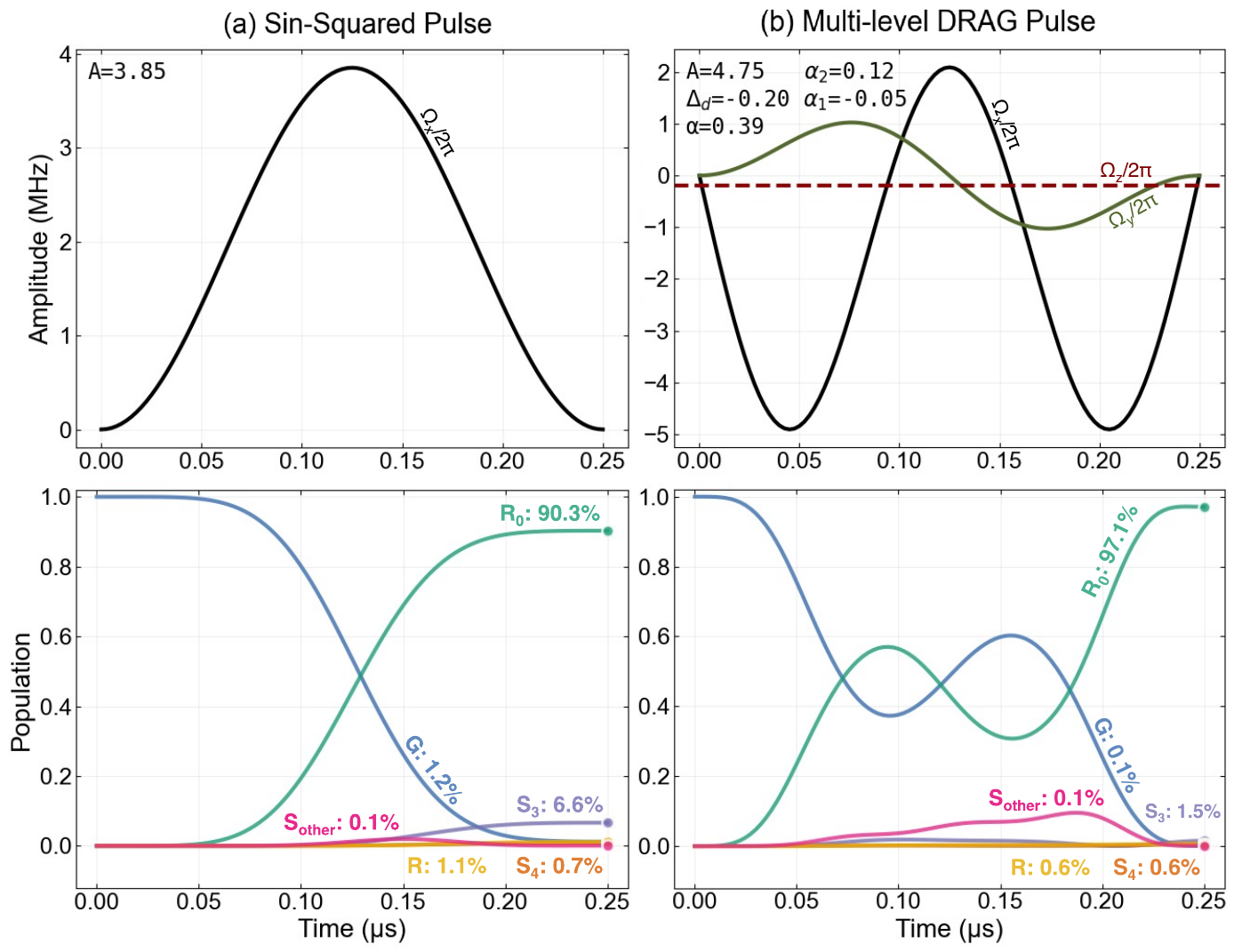}}
    \caption{\textbf{Performance of DRAG pulses in the dephasing-free regime.} The top row shows (a) a sine-squared pulse and (b) a multi-level perturbative DRAG pulse, where $\Omega_x$ denotes the in-phase primary drive (black), $\Omega_y$ the out-of-phase counterdiabatic drive (dark green), and $\Omega_z$ the constant laser detuning (dark red). The bottom row shows the corresponding dynamics for $T = 0.25~\mu\mathrm{s}$ and $\sigma = 4.6~\mu\mathrm{m}$. Populations in the relevant states are shown: ground state $G$ (blue), singly-excited symmetric state $\ket{R_0}$ (green), all singly-excited asymmetric states $R$ (yellow), leakage dominated doubly-excited states $S_3$ and $S_4$ (purple and orange), and remaining doubly-excited states $S_{\mathrm{other}}$ (pink). Populations in the continua $M$ and $M_{\mathrm{th}}$ are zero in this regime. Final populations are highlighted with matching colors. In the absence of Markovian dephasing, the DRAG pulse effectively suppresses coherent leakage into the dominant channels, increasing the population in $\ket{R_0}$ from $90.3\%$ to $97.1\%$.}
    \label{fig:Initial_parameter_plots_withoutdephasing_optimized}
\end{figure}

As discussed in Sec.~\ref{sec:expt_dynamics}, we analyze the dephasing-free regime here to identify the fundamental performance limits of the system. This analysis demonstrates that DRAG pulse shaping effectively suppresses unitary leakage in the absence of non-unitary dephasing, showing that the maximal performance achievable in the full model is ultimately limited by Markovian dephasing rather than by the control strategy itself. By ``dephasing-free regime,'' we refer to the case in which irreversible decay into both the interaction-induced continuum $M$ and the thermal continuum $M_{\mathrm{th}}$ is excluded from the dynamics, while all coherent couplings to the asymmetric singly- and doubly-excited states explicitly included in the system Hamiltonian are retained. From this perspective, a dephasing-free regime does not correspond to a physically realizable experimental limit, since interaction-induced dephasing cannot be completely eliminated. Thermal dephasing, in contrast, can in principle be reduced through improved laser performance and lower atomic temperatures. In this section, however, we neglect both thermal and interaction-induced Markovian dephasing purely as a theoretical tool, allowing us to isolate and analyze the coherent control dynamics.

In the main text, we introduced a single-level perturbative DRAG pulse designed to suppress leakage into a single dominant leakage channel [see Eq.~(\ref{eq:Perturbative_drag_1})]. Here, we extend this approach to a multi-level perturbative DRAG pulse. As shown in Sec.~\ref{sec_results}, the dominant leakage channels correspond to the states $S_3$ and $S_4$, with $S_3$ exhibiting the largest population followed by $S_4$ for the experimental parameters [see Fig.~\ref{fig:leakage}]. This motivates the use of a multi-level DRAG pulse that incorporates counterdiabatic corrections associated with both leakage channels, thereby suppressing population transfer to $S_3$ and $S_4$ simultaneously. The resulting pulse is given by
\begin{equation}
\begin{split}
\Omega_{\text{DRAG}}^{\text{(multi-pert)}}(t) = & \Bigg[ \Omega_I(t) 
- i \Big(\frac{\alpha}{\delta_{S_3}} + \frac{\alpha_1}{\delta_{S_4}}\Big) \dot{\Omega}_I(t) \\
& + \frac{\alpha_2 \ddot{\Omega}_I(t)}{\delta_{S_3} \delta_{S_4}} \Bigg] e^{i \Delta_d t},
\end{split}
\label{eq:multi-level_Perturbative_drag_1}
\end{equation}
where $\delta_{S_3}$ and $\delta_{S_4}$ denote the energy gaps between the respective leakage states and the target state $\ket{R_0}$. The pulse contains five optimization parameters: the amplitude $A$ of the primary pulse, the constant laser detuning $\Delta_d$, and three DRAG coefficients $\alpha$, $\alpha_1$, and $\alpha_2$, which control the strength of the counterdiabatic corrections. These parameters are optimized using the same procedure described in the main text, with the objective of minimizing leakage from the target subspace.

The results of the optimization are shown in Fig.~\ref{fig:Initial_parameter_plots_withoutdephasing_optimized}. In the dephasing-free regime, with experimental parameters \(T = 0.25~\mu\mathrm{s}\) and \(\sigma = 4.6~\mu\mathrm{m}\), the population in the target state \(\ket{R_0}\) increases from \(90.3\%\) for a simple sine-squared pulse to \(97.1\%\) with the optimized multi-level DRAG pulse. In the absence of Markovian dephasing, the DRAG protocol leverages other doubly-excited states to recover population from $\delta_{S_3}$ and $\delta_{S_4}$ back into the target state. This substantial improvement demonstrates that DRAG control is highly effective at suppressing unitary leakage. In contrast, when rapid non-unitary dephasing is present, population can irreversibly leak before the corrective action of the DRAG pulse can take effect, preventing full recovery to the target state, as seen in Fig.~\ref{fig:Initial_parameter_plots_dephasing_optimized}. This also explains why single-level and multi-level DRAG pulses exhibit similar performance in the presence of dephasing. For example, when Markovian dephasing is included, the optimized multi-level DRAG pulse for the experimental parameters achieves a population of \(87.7\%\) in \(\ket{R_0}\), comparable to that obtained with the single-level DRAG protocol [see Fig.~\ref{fig:Initial_parameter_plots_dephasing_optimized}]. This justifies the use of the simpler single-level DRAG pulse in the main text. We note that optimizations performed using dCRAB yield lower fidelities in this regime. Overall, the dephasing-free analysis confirms that strong Markovian dephasing, rather than limitations of the control protocol itself, constitutes the fundamental bottleneck to achieving higher target state transfer fidelities in the system.

\section{Model Parameters}
\label{app_leakage}
\setcounter{figure}{0}

We provide here the full set of parameters used in the simulations in the main text. For $n_{\text{max}} = 8$, the Rydberg-Rydberg interaction shifts of the doubly-excited states are given by
\begin{widetext}
\begin{equation}
\{\delta_{S_0}, \delta_{S_1}, \delta_{S_2}, \delta_{S_3}, \delta_{S_4}, \delta_{S_5}, \delta_{S_6}, \delta_{S_7}\}
= -\left( \frac{C_6}{\sigma^6} \right) \times
\{796.041,41.959,5.877,1.214,0.320,0.099,0.033,0.011\}
\times 10^{-4},
\label{eq:energy_shifts}
\end{equation}
with the corresponding decay rates into the continuum $M$ given by
\begin{equation}
\{\gamma_{S_0}, \gamma_{S_1}, \gamma_{S_2}, \gamma_{S_3}, \gamma_{S_4}, \gamma_{S_5}, \gamma_{S_6}, \gamma_{S_7}\}
= \left( \frac{C_6}{\sigma^6} \right) \times
\{1383.691,50.946,4.659,0.709,0.150,0.031,0.012,0.004\}
\times 10^{-4}.
\label{eq:decay_rates}
\end{equation}
Here, $C_6 = 2\pi \times 154~\text{THz}~\mu\text{m}^6$ is the van der Waals interaction coefficient for the chosen Rydberg state $\ket{r} = \ket{109S_{1/2}, J = 1/2, m_J = 1/2}$, and $\sigma$ denotes the radius of the atomic ensemble. The coupling coefficient matrix $\beta_{jk}$ is given by
\begin{equation}
\beta_{jk} =
\renewcommand{\arraystretch}{1.4}
\setlength{\arraycolsep}{1.8pt}
\left(
\begin{array}{c|cccccccc}
& \textbf{S}_0 & \textbf{S}_1 & \textbf{S}_2 & \textbf{S}_3 & \textbf{S}_4 & \textbf{S}_5 & \textbf{S}_6 & \textbf{S}_7 \\ \hline
 \textbf{R}_0 & 6.6400 & -9.0300 &  7.1500 &  3.5200 &  1.1400 &  0.2300 & -0.0248 &  0.0010 \\
 \textbf{R}_1 & 3.0900 & -1.3700 & -2.8000 & -4.3600 & -2.7100 & -0.8910 &  0.1460 & -0.0086 \\
 \textbf{R}_2 & 1.5400 & -0.8710 &  0.8490 &  2.6900 &  3.3000 &  1.8100 & -0.4580 &  0.0407 \\
 \textbf{R}_3 & 0.7420 & -0.6850 &  1.4500 &  1.6400 & -0.1800 & -1.3100 &  0.7010 & -0.1080 \\
 \textbf{R}_4 & 0.3490 & -0.5250 &  1.0400 &  0.4410 & -0.3620 &  0.2900 & -0.6340 &  0.1900 \\
 \textbf{R}_5 & 0.1620 & -0.3720 &  0.5580 &  0.0421 &  0.0389 &  0.2160 &  0.2510 & -0.2280 \\
 \textbf{R}_6 & 0.0738 & -0.2460 &  0.2510 &  0.0045 &  0.1170 & -0.0424 &  0.0294 &  0.1670 \\
 \textbf{R}_7 & 0.0332 & -0.1540 &  0.0947 &  0.0330 &  0.0479 & -0.0420 & -0.0560 & -0.0640 \\
\textbf{R}_8 & -0.0275 & -0.0237 & -0.0203 &  0.0173 & -0.0147 &  0.0123 &  0.0102 &  0.0082
    \end{array}
\right) \times 10^{-1}.
\label{eq:coupling_coefficient}
\end{equation}
\end{widetext}
In particular, the first row of this matrix corresponds to the couplings $\beta_{0k}$ between the target state $\ket{R_0}$ and the doubly-excited states $\ket{S_k}$.

\section{Simulation and Optimization Methods}
\label{app_optimization}
\setcounter{figure}{0}

We provide here detailed information on the computational tools used to simulate the time evolution of the system, optimize the DRAG pulses, and perform GRAPE-based pulse optimizations.

The time evolution under various control pulses is simulated using the QuTiP library~\cite{lambert_2024_qutip_arxiv}, specifically the \texttt{mesolve} function. Starting from the initial state $\ket{G}$, \texttt{mesolve} numerically integrates the Lindblad master equation [see Eq.~\eqref{eq:Master_eq}], which governs the system dynamics. Dissipative processes are incorporated via the collapse operators (\texttt{c\_ops}), while the populations of the singly- and doubly-excited states, as well as the associated continua, are monitored through expectation operators (\texttt{e\_ops}). The evolution is computed on a discrete time grid of 100 points, and the solver is allowed up to 50000 internal integration steps (\texttt{nsteps}) to ensure numerical stability and accuracy in the large Hilbert space.

For DRAG pulse optimization, the loss function defined in Eq.~(\ref{Eq:Loss_function}) is minimized in order to maximize the population transfer to the target state $\ket{R_0}$. The optimization is performed over three control parameters: the pulse amplitude $A$, the DRAG coefficient $\alpha$, and the constant laser detuning $\Delta_d$. We adopt a two-step numerical optimization strategy. In the first step, a global search is carried out using the \texttt{differential\_evolution} algorithm implemented in the SciPy library~\cite{Pauli_2020_SciPy_NMeth}. This approach explores a broad parameter space and mitigates convergence to sub-optimal local minima, providing a robust initial estimate for subsequent refinement. The parameter ranges are constrained within the experimentally feasible values: $A$ and $\alpha$ are restricted to $2\pi \times [-3/T,\, 3/T]~\mathrm{MHz}$, and $\Delta_d$ is limited to $2\pi \times [-5,\, 5]~\mathrm{MHz}$. In the second step, the parameters obtained from the global search are refined using the L-BFGS-B algorithm~\cite{Liu_MathProg_1989}, which performs a gradient-based local optimization subject to the same bound constraints. This procedure yields precise and stable convergence to the minimum of the loss function. The combination of global and local optimization ensures comprehensive exploration of the parameter space followed by reliable convergence to the optimal DRAG pulse parameters. We verified that the optimized pulse parameters are robust against variations in time discretization and solver tolerances.

To benchmark the performance of DRAG control pulses, GRAPE optimizations are performed using the \texttt{minimize} function from the SciPy library. The control fields \( \Omega_x(t) \) (primary control field), \( \Omega_y(t) \) (counterdiabatic control field), and \( \Omega_z(t) \) (laser detuning) are discretized into \( n = 50 \) time steps for each parameter set, with initial amplitudes seeded from the previously optimized DRAG pulse shapes. We verified convergence by increasing the number of time steps up to \( n = 100 \), yielding identical fidelities. The final amplitudes are constrained within \( 2\pi \times [-3/T, \, 3/T]~\mathrm{MHz} \), providing sufficient flexibility to explore the control landscape. Additional constraints ensure physically realizable pulses by enforcing that the amplitude starts and ends at zero and that the maximum change between consecutive time steps does not exceed \( 0.5~\mathrm{MHz} \). The optimization is carried out using the SLSQP algorithm, which incorporates these nonlinear constraints alongside the amplitude bounds. The same loss function as in the DRAG optimization is employed, targeting maximal population in the target Rydberg state \( \ket{R_0} \). To obtain smooth, continuous control pulses suitable for experimental implementation, the optimized discrete GRAPE pulses are processed using the Savitzky-Golay filter from the \texttt{scipy.signal} library, with a window length of 7 and a polynomial order of 3. This procedure effectively fits a local polynomial of order 3 to consecutive segments of the discrete pulse with a window length of 7, suppressing high frequency fluctuations while preserving the overall pulse shape and key features. The fidelities were recomputed after filtering, and no significant degradation was observed.

\bibliographystyle{apsrev4-2}
\bibliography{ref_superatom}

\end{document}